\pdfoutput=1
\documentclass[journal,onecolumn,12pt]{IEEEtran}

\usepackage{graphicx}
\usepackage{tabularx}                   
\usepackage{longtable}                  
\newcommand{\specialcell}[2][c]{%
  \begin{tabular}[#1]{@{}c@{}}#2\end{tabular}} 
\usepackage{xcolor,colortbl}            
\usepackage{setspace}
\usepackage{epstopdf}  
\usepackage{float}                      
\usepackage{tikz}                       
\usepackage{lipsum}
\usepackage{algorithm}
\usepackage[noend]{algpseudocode}
\makeatletter
\algrenewcommand\ALG@beginalgorithmic{}
\makeatother

\usepackage{svg}
\usepackage{amssymb}
\usepackage{amsmath}
\usepackage{xspace}
\usepackage{url}
\usepackage[shortcuts,acronym]{glossaries}
\newacronym{sdn}{SDN}{Software-Defined Network}
\newacronym{fgkkf}{FKKF}{Frequency-based Kernel Kalman Filter}
\newacronym{tcam}{TCAM}{Ternary Content-Addressable Memory}
\newacronym{pcie}{PCIe}{Peripheral Component Interconnect Express}
\newacronym{kkr}{KKR}{Kernel Kalman Rule}
\newacronym{gkkf}{GKKF}{Generalized Kernel Kalman Filter}
\newacronym{ip}{IP}{Internet Protocol}
\newacronym{tcp}{TCP}{Transmission Control Protocol}
\newacronym{cpu}{CPU}{Central Processing Unit}
\newacronym{asic}{ASIC}{Application Specific Integrated Circuit}
\newacronym{mft}{MFT}{Multiple Flow Tables}
\newacronym{vlan}{VLAN}{Virtual Local Area Network}
\newacronym{qos}{QoS}{Quality of Service}
\newacronym{hmm}{HMM}{Hidden Markov Model}
\newacronym{kf}{KF}{Kalman Filter}
\newacronym{em}{EM}{Expectation-Maximization}
\newacronym{ekf}{EKF}{Extended Kalman Filter}
\newacronym{ukf}{UKF}{Unscented Kalman Filter}
\newacronym{rkhs}{RKHS}{Reproducing Kernel Hilbert Space}
\newacronym{rbf}{RBF}{Radial Basis Function}
\newacronym{ditg}{D-ITG}{Distributed Internet Traffic Generator}
\newacronym{ovs}{OVS}{Open vSwitch}
\newacronym{vm}{VM}{Virtual Machine}
\newacronym{idt}{IDT}{Inter Departure Time}
\newacronym{kvm}{KVM}{Kernel-based Virtual Machine}
\newacronym{ovsdb}{OVSDB}{Open vSwitch Database Management Protocol}
\newacronym{arp}{ARP}{Address Resolution Protocol}
\newacronym{http}{HTTP}{Hypertext Transfer Protocol}
\newacronym{api}{API}{Application Programming Interface}
\newacronym{gre}{GRE}{Generic Routing Encapsulation}
\newacronym{vnc}{VNC}{Virtual Network Computing}
\newacronym{udp}{UDP}{User Datagram Protocol}
\newacronym{cli}{CLI}{Command-Line Interface}
\newacronym{json}{JSON}{JavaScript Object Notation}
\newacronym{os}{OS}{operating system}
\newacronym{pca}{PCA}{Principal Component Analysis}
\newacronym{te}{TE}{traffic engineering}
\newacronym{ecmp}{ECMP}{Equal-Cost Multi-Path}
\newacronym{tor}{ToR}{Top of Rack}
\newacronym{cre}{CRE}{Cognitive Routing Engine}
\newacronym{cram}{CRAM}{Cognitive Routing Algorithm Module}
\newacronym{rnn}{RNN}{Random Neural Network}
\newacronym{nmm}{NMM}{Network Monitoring Module}
\newacronym{kkf}{KKF}{Kernel Kalman Filter}
\newacronym{kkf-ceo}{KKF-CEO}{Kernel Kalman Filter based on the Conditional Embedding Operator}
\newacronym{arma}{ARMA}{Autoregressive Moving Average}
\newacronym{arima}{ARIMA}{Autoregressive Integrated Moving Average}
\newacronym{garch}{GARCH}{Generalized Autoregressive Conditional Heteroskedasticity}
\newacronym{blrnn}{BLRNN}{Bilinear Recurrent Neural Network}
\newacronym{mlp}{MLP}{Multilayer Perceptron}
\newacronym{anfis}{ANFIS}{Adaptive Neuro-Fuzzy Inference System}
\newacronym{svm}{SVM}{Support Vector Machine}
\newacronym{snmp}{SNMP}{Simple Network Management Protocol}
\newacronym{ft}{FT}{Fourier Transform}
\newacronym{dft}{DFT}{Discrete Fourier Transform}
\newacronym{stft}{STFT}{Short-Time Fourier Transform}
\newacronym{fft}{FFT}{Fast Fourier Transform}
\newacronym{pc}{PC}{principal component}
\newacronym{wan}{WAN}{Wide Area Network}
\newacronym{onl}{ONL}{Open Network Linux}
\newacronym{dc}{DC}{Datacenter}

\definecolor{tbl_g}{RGB}{172, 239, 182}
\definecolor{tbl_y}{RGB}{239, 236, 172}
\definecolor{tbl_r}{RGB}{239, 172, 172}

\newcommand{\pjnote}[1]{}
\newcommand{\gnnote}[1]{}

\newcommand{\secs}{s\xspace}
\newcommand{\msecs}{ms\xspace}

\newcommand{\indiv}{\textsc{[indiv]}}
\newcommand{\other}{\textsc{[inter]}}
\newcommand{\nonlin}{\textsc{[nonlin]}}
\newcommand{\simult}{\textsc{[scale]}}

\usepackage{authblk}
\usepackage{enumitem}

\begin{document}
\title{Towards Fine Grained Network Flow Prediction}

\begin{spacing}{1.7}

\author{
  \IEEEauthorblockN{
    Patrick Jahnke\IEEEauthorrefmark{1}, 
    Emmanuel Stapf\IEEEauthorrefmark{1}, 
    Jonas Mieseler\IEEEauthorrefmark{1}, 
    Gerhard Neumann\IEEEauthorrefmark{1}\IEEEauthorrefmark{3}, 
    Patrick~Eugster\IEEEauthorrefmark{1}\IEEEauthorrefmark{2}\\}
  \IEEEauthorblockA{
    \IEEEauthorrefmark{1}Department of Computer Science, TU Darmstadt\\
    }
      \IEEEauthorblockA{
    \IEEEauthorrefmark{2}Faculty of Informatics, Universit\`{a} della Svizzera italiana\\
    }
      \IEEEauthorblockA{
    \IEEEauthorrefmark{3}School of Computer Science, University of Lincoln\\
    }
}


\maketitle

\begin{abstract}
One main challenge for the design of networks is that traffic load is not generally known in advance.
This makes it hard to adequately devote resources such as to best prevent or mitigate bottlenecks.
While several authors have shown how to predict traffic in a coarse grained manner by aggregating flows, fine grained prediction of traffic at the level of individual flows, including bursty traffic, is widely considered to be impossible. 
This paper shows, to the best of our knowledge, the first approach to fine grained per-flow traffic prediction. In short, we introduce the \ac{fgkkf}, which predicts individual flows' behavior based on measurements. 
Our \ac{fgkkf} relies on the well known Kalman Filter in combination with a kernel to support the prediction of non linear functions. 
Furthermore we change the operating space from time to frequency space. In this space, into which we transform the input data via a \ac{stft}, 
the peak structures of flows can be predicted after gleaning their key characteristics, with a \ac{pca}, from past and ongoing flows that stem from the same socket-to-socket connection.
We demonstrate the effectiveness of our approach on popular benchmark traces from a university data center. 
Our approach predicts traffic on average across 17 out of 20 groups of flows
with an average prediction error of 6.43\% around 0.49 (average) seconds in advance, whilst 
existing coarse grained approaches exhibit prediction errors of 77\% at best.
\end{abstract}


%
\IEEEpeerreviewmaketitle

\section{Introduction}
\label{sec:introduction}
\IEEEPARstart{T}{oday's} data centers execute a variety of applications and services, 
with increasing network traffic loads.
One typical problem that arises in networks that have to deal with large amounts of traffic is \emph{congestion} -- 
when a network device is receiving more data packets than it can process, packets are delayed or even dropped, 
inevitably lowering performance of applications and services. 
The reason for congestion can be found in the \emph{bursty} nature of certain network traffic with multiple traffic flows transmitted on the same link producing high peaks simultaneously. 
By increasing network bandwidth, 
nowadays up to 100Gb/s, 
the problem is not immediately solved, as many so-called elephant flows~\cite{elephant} in low bandwidth networks in fact result from peaks which are ``flattened''; in high bandwidth networks these retain their original (bursty) nature.

Similarly, there is a high chance that flow completion times are not decreased with increased bandwidth in the presence of congestion when relying on reactive congestion control mechanisms (e.g., TCP, DCTCP~\cite{dctcp}, PCC~\cite{pcc}, RCP~\cite{rcp}, or XCP~\cite{xcp}), since these incur further communication between sender and receiver and thus delays in reacting to congestion~\cite{Jose:2015:HSN:2834050.2834096}.
Therefore new congestion control algorithms are needed for a better bandwidth utilization~\cite{alizadehattar_et_al:DR:2016:6760},~\cite{Jose:2015:HSN:2834050.2834096}.
Ideally a network would be capable of \emph{predicting} the evolution of traffic flows so that an impending over-utilization would be recognized early enough to appropriately re-assign flows to other paths before real packet loss or congestion occur.

Virtually all previous research on network traffic prediction however considers traffic data which is \emph{aggregated in time and space}, i.e.,  
considering flows over long periods of time, 
combining many such flows, 
and predicting for the same (aggregated) flows. 
Since most congestion is caused by the interaction of short traffic bursts from elephant flows, 
these \emph{coarse grained} prediction methods cannot 
usefully predict such bursts in \emph{individual} traffic flows. 
In fact, 
\emph{fine grained} per-flow prediction has thus far been widely considered to be infeasible~\cite{microte}.
More precisely, 
in the context of this paper we are interested in prediction 
\begin{description}
\item[\bf\indiv]~ at the level of \emph{individual} flows to  enable adequate adaptations especially under congestion; 
\item[\bf\other]~~ for flows from the past of \emph{other} flows and not only the very same flows (inter- vs intra-flow learning), to avoid delayed or missing responses due to limited information being available;
\item[\bf\nonlin]~~\; capturing \emph{non-linear behavior} as flows can exhibit high variance in short time, a key characteristic also of bursts;
\item[\bf\simult]~~ of large numbers of flows simultaneously to \emph{scale} to entire networks.  
\end{description}

Needless to say that such prediction is only useful if it happens sufficiently \emph{ahead of time} to enable reaction, and achieves \emph{high accuracy}. 

This paper presents, to the best of our knowledge, the first approach to such fine grained prediction.
We view a flow in a network as a non-linear system. 
The traffic load can be seen as the observations, with flows constituting time series of kbit values. Several latent factors contribute to the true very complex \emph{hidden} state of the system, e.g., type of programs engendering the flows and their workloads, user behavior,  drivers running on the respective hosts, 
network elements like interface cards and switches forwarding the traffic or the links between them. 

Many network flows typically show a very bursty behavior and predicting these bursts in the time domain seems infeasible due to the high frequency and stochasticity of such flows. 
We observe that many flows show a more distinctive pattern in  the \emph{frequency domain}. 
The underlying assumption of our approach is that the evolution of these frequency patterns can be predicted. 
However, in the frequency domain the observations correspond to \acfp{ft} of observation time windows,  constituting a high-dimensional observation space. 
Hence, traffic flow prediction requires prediction methods that can deal with hidden states, non-linear system dynamics and high dimensional observations.

Given these requirements, we introduce a key novel prediction technique dubbed the \acf{fgkkf}, which uses the the powerful concept of \acp{hmm} for modeling the system. In short, our approach is inspired by the recently introduced \acf{kkr}~\cite{KKR}, yet operates in the frequency space and together with a \acf{pca}. 
In the frequency space, into which we get input data via \ac{stft}, 
the non-linear peak structures of unseen flows can be predicted efficiently after gleaning their key characteristics, with a \ac{pca}, from past and ongoing flows that stem from the same socket-to-socket connection. 

Our evaluation results demonstrate the effectiveness of our approach. 
In a primary step, 
popular real-world benchmark traces from a university data center were analyzed in order to find structures in the diverse flow set that could be learned.  
Eventually, repeating structures were found for flows stemming from recurring socket-to-socket connections.  
Subsequently, the selected traffic data was used in traffic prediction experiments, employing our \ac{fgkkf} for dealing with the learning and prediction problem.  
In short our approach predicts traffic across 17 out of 20 groups of flows with an average prediction error of 6.43\%, and around 0.49 (average) seconds in advance which we believe is sufficient for many \ac{te} approaches~\cite{google, swan, te_sdn_2013, hedera, mahout}.  

In summary, this paper makes the following concrete contributions:

\begin{itemize}
    \item A novel approach for \emph{fine grained} prediction of the trajectory of individual (and yet unseen) flows. 
    \item An algorithm as well as system design for implementing our approach on \emph{commodity} \ac{dc} switches.
    \item An evaluation of our approach showing its high accuracy as well as scalability. We also include a comparison against existing coarse grained time-series approaches, showing how these  yield prediction errors beyond 77\%.
\end{itemize}

The rest of the paper is structured as follows. 
Section~\ref{sec:relatedWork} introduces related prior work.
Using foundations of time series modeling and \ac{rkhs} the formulations of the \ac{fgkkf} are derived in Section~\ref{sec:gkkf}. 
Our algorithm for implementing our learning and prediction approach and its implementation are presented in Section~\ref{sec:gkkf_algo}. 
The results of the traffic prediction experiments using the \ac{fgkkf} are shown in Section~\ref{sec:evaluation}.
Finally we draw conclusions and discuss future work in Section~\ref{sec:conclusion}.

\section{Related Work}
\label{sec:relatedWork}

\subsection{Approaches in Network Flow Prediction} 
The general topic of network traffic prediction was already subject of extensive research~\cite{forcast}. 
One of the simplest approaches to model a time series of traffic data is the \gls{arma} model which consists of an autoregressive part performing a regression on the series of data points and a moving average part that tries to model the error of the time series. 
In the literature \gls{arma} was used to predict network traffic mostly from single applications like BitTorrent~\cite{arma_bitt} or FTP~\cite{arma_ftp}. 
Since \gls{arma} is only applicable for time series produced by stationary stochastic processes, the \gls{arima} model was subsequently developed. 
\gls{arima} and variants of it were used in different scenarios for traffic prediction, e.g., in 3G mobile networks~\cite{arima_3g} or public safety networks~\cite{arima_psn}.
Like \gls{arma}, \gls{arima} is only applicable for linear time series with constant variance (cf. \nonlin).  
The \gls{garch} model~\cite{garch} was introduced to model non-linear time series with a time dependent variance. 
The authors show that the model performs better than \gls{arima} in capturing the bursty nature of Internet traffic whose variance changes over time.  

Neural networks similarly can model non-linear time series, which is why already many different types of neural networks  were tested for predicting network traffic.  
Park and Woo~\cite{d-blrnn} for instance apply dynamic \glspl{blrnn}, showing superior performance compared to static \glspl{blrnn} or classical neural networks like the \gls{mlp} previously also used for prediction~\cite{mlp,nn}.  
Other works focus on combining neural networks with linear approaches like \gls{arima} under the assumption that the traffic time series consists of linear and non-linear components~\cite{nn_arima}. 
Yet other approaches connect neural networks with other methodologies, e.g., with fuzzy systems in the \gls{anfis}~\cite{anfis}. 
\glspl{svm} were also used for predicting network traffic, e.g., by Liang et al.~\cite{svm_ant}, by selecting parameters using an ant colony optimization algorithm.
A generic algorithm is used by ~\cite{tcp-ga} which attempts to find the best matching combination of mathematical functions that approximate a given time series that accounts for TCP throughput. 
However, none of the above or other approaches are fine grained -- the used \emph{data was always aggregated} in two ways (cf. \indiv):

\paragraph{Time}
The data sets consisted of traffic data collected during a time span of days, weeks, or even months, 
which was aggregated \emph{on a temporal scale} leading to sampling intervals for the traffic load between 1 second and 1 hour. 
Only in~\cite{neural_tree} an interval of 0.1 seconds was used. 
By aggregating the traffic data, time series become less complex and much easier to learn. 
However, since the characteristics of individual flows are not present in the data set anymore, they cannot be predicted either. 
This is a problem because already rather short traffic peaks with very high loads can cause congestion in a network ~\cite{traffic_char}. In order to predict \emph{individual} flows small sampling intervals are needed as otherwise short high-volume flows would be represented by only a few data points (see Figure~\ref{fig:flow1s}).

\paragraph{Space}
The selected data was also aggregated \emph{in terms of flows}, 
meaning that all socket-to-socket connections either between same hosts or same network elements were combined to one big flow, 
even across different respective protocols (e.g. TCP and UDP).
Again, 
this hides the bursty characteristics of the traffic and leads to the assumption that peaks are very rare. 
As a result, 
a lot of helpful applications of different systems (e.g., \ac{te} systems as shown in~\cite{microte}) are not able to identify and predict single high-volume flows in the network in a fine-grained manner. 

In addition, in all the aforementioned approaches flows were split into training and testing parts. 
The prediction model is thus entirely learned from the very flow during its life time which is very difficult for short high-volume flows (cf. \other).

\subsection{State Estimation of Non-linear Systems}
Approaches for non-linear time series modeling that are related to our \gls{fgkkf} are the \gls{kkf}~\cite{kkf} and the \gls{kkf-ceo}~\cite{kkf-ceo}. 
The \gls{kkf} is a variant of the \gls{kf} where the observations, 
system states, and update equations of the \gls{kf} are brought to a feature space by using a kernel function.  
However,  in contrast to our \gls{fgkkf}, a kernelized version of the system model is only derived for the transition model but not for the observation model.
As a result, the observations are computed from the states simply by adding a noise term that can lead to wrong assumptions about the prediction accuracy. 
Another difference to the \gls{fgkkf} is that \gls{kf} formulations are only embedded in a \emph{sub-space} of the feature space and hence, 
the approach is not fully exploiting the infinite-dimensionality characteristic of the feature space.
The \gls{kkf-ceo} embeds the formulations of the \gls{kf} in a \gls{rkhs} by using the conditional embedding operator which is explained in more detail along with the \gls{rkhs} in Section~\ref{sec:rkhskf}. 
In contrast to the \gls{kkf}, \gls{kkf-ceo} formulates the \gls{kf} in the full feature space provided by the kernel.  
Moreover, the transition model does not have to be learned using the \gls{em} algorithm, as with the \gls{kkf}, but can be computed from the training data. 
However, as for the \gls{kkf}, the observation model is not formulated in the \gls{rkhs} and the observations again are interpreted as noisy variants of the system states. 
Computing the transition model under this assumption using the embeddings of the noisy observations is not fully valid from a theoretical standpoint since the observations were not generated by a Markov  process.
Furthermore, this leads to update equations that are further away from the original \gls{kf} equations compared to our \gls{fgkkf}. 
\section{\acl{fgkkf}}
\label{sec:gkkf}

\begin{figure}
\centering
\includegraphics[width=18cm]{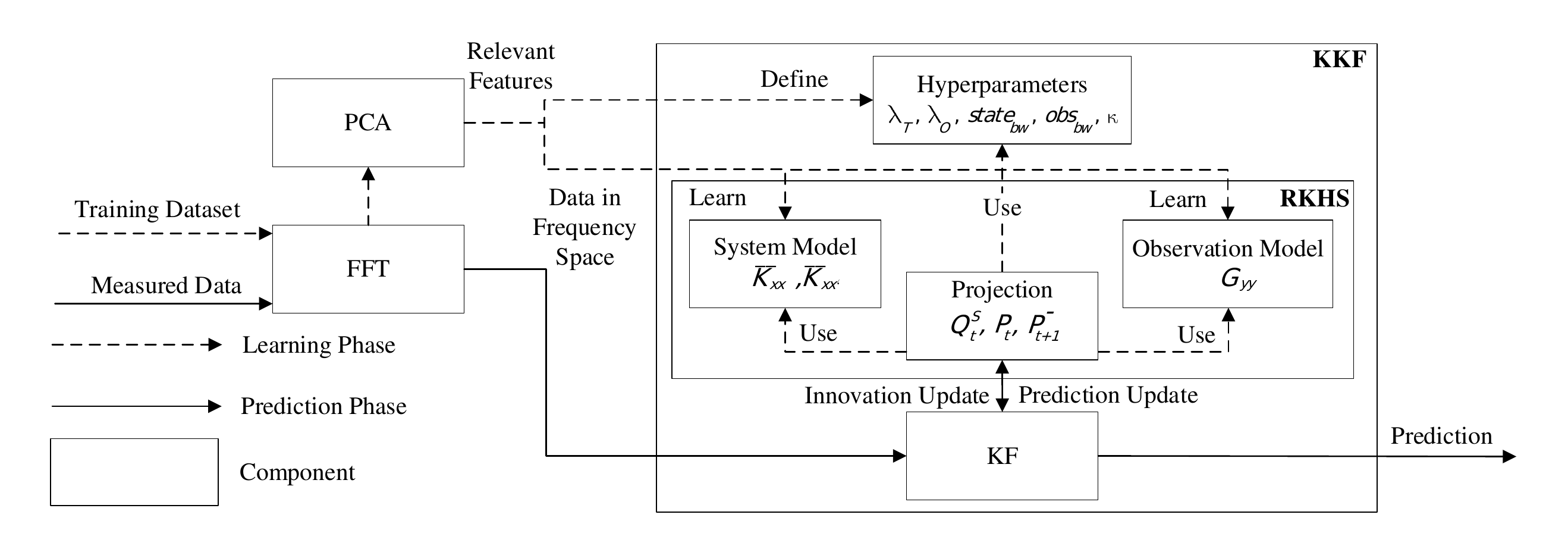}
\caption{Overview of the FKKF}
\label{fig:overview}
\end{figure}
\label{sec:traffic_pred}

As mentioned in Section~\ref{sec:introduction}, a traffic flow can be seen as a non-linear system with hidden states, unknown dynamics and high-dimensional observations in frequency domain. 
The \acf{fgkkf} is tailored for prediction problems in non-linear systems with hidden states and high-dimensional observations in the frequency domain as suitable fine grained for flow prediction.

\subsection{Overview}
The \ac{fgkkf} uses Kalman filtering as efficient inference technique for state estimation and further prediction~\cite{KKR}. 
However, as opposed to standard Kalman filtering, which only works in linear systems, the \ac{fgkkf} uses a hidden state representation that is embedded in a \acf{rkhs}. 
This space represents a non-linear transformation of the original state space into a possibly infinite dimensional kernel space. 
While the system and observation models for the systems of interest are typically non-linear in the original state space, they can be approximated efficiently in the \ac{rkhs} by linear models~\cite{KKR}. 
Hence, standard Kalman filtering can be applied to this \ac{rkhs} representation of the system. In addition 
the \ac{fgkkf} transforms its observations into the frequency domain in order to cope with the challenges of flow prediction.

Figure~\ref{fig:overview} gives an overview of our \ac{fgkkf}. 
It combines --- and inherits from --- several  known techniques, which are \ac{kf}, \ac{rkhs}, \ac{fft}, and \ac{pca}. 
The dashed lines show the learning phase, where relevant features are extracted from the training data set. 
After transforming the features into frequency space, a system model and observation model are learned. 
Several hyperparameters (see Table~\ref{table:paraFGKKF}) are defined to regularize the system and observation model, and to scale the bandwidth of the kernel function.
The solid lines show the prediction phase where the measured data are also transformed into frequency space.
These data are used for the innovation and prediction update of the Kalman filter. Since the system and observation model are linear in the \ac{rkhs}, a projection between the \ac{rkhs} and the original space is needed. 
This projection uses the learned models and defined parameters.

The \ac{fft} is a well known algorithm to transform a signal from the time domain into the frequency space~\cite{FFT}. 
For the flow prediction application, we have observed that prediction in frequency space is simpler than in the time domain as bursty traffic often looks arbitrary in time space but has a noticeable characteristic, which can be predicted, in the frequency space (see Figure~\ref{fig:comp_tf}). 
We use \ac{pca} to reduce the dimensionality of our observation  space with a minimal amount of information loss~\cite{jolliffe2002principal}. 
We also use an extension of the \ac{kkf} formulation called the sub-space \ac{kkf}, which uses a sparse representation of the \ac{rkhs} to facilitate computation on large data-sets. 
Next we give more details of our \gls{fgkkf}.

\subsection{Kernel based Kalman Filter}
We assume to receive an observation $y_t$ from the system at time point $t$ whose true hidden state is denoted by $x_t$. 
We represent our belief over state $x_t$ with a mean embedding $\vec \mu_{\varphi(x_t)}$ in a \gls{rkhs} 
with the kernel $k(\vec x_t,\vec x_{t+1}) = \langle \vec \varphi(\vec x_t),\, \vec \varphi(\vec x_{t+1}) \rangle$ 
and with a covariance embedding $\vec \Sigma_{\vec \varphi(\vec x_t)}$ in a tensor-product \gls{rkhs} with the kernel 
$h(\vec x_t,\vec x_{t+1}) = \langle \hat{\vec \varphi}(\vec x_t),\, \hat{\vec  \varphi}(\vec  x_{t+1}) \rangle$, where $\hat{\vec \varphi}(\vec x_t) = \vec \varphi(\vec x_t) - \vec{\mu}_{\varphi(x_t)}$ is the centered feature mapping.
The mean embedding corresponds to the mean of the Gaussian belief distribution while the covariance embedding corresponds to the covariance in standard Kalman filtering. 
Moreover, the distribution over the observation $y_t$ is embedded into another \gls{rkhs}. For easier distinction, this feature mapping is denoted by $\phi(y_t)$. 
We write $\vec \varphi_t$ and $\vec \varphi_{t+1}$  instead of $\vec \varphi(\vec x_{t})$ and $\vec \varphi(\vec x_{t+1})$ respectively for simplicity.

The Kalman filter equations are composed of two updates -- ($i$) the prediction update, which maps the current belief state to the next time step, and ($ii$) the innovation update, which incorporates
the current observation in our belief state. As our system dynamics model as well as the observation model are unknown, both quantities need to be estimated from data. In the remainder
of this section, we assume that we have access to a training data set, containng the state and observation trajectories, $\vec x_{1:T}$ to $\vec y_{1:T}$. 
The goal is to devise a state estimation and prediction algorithm that also works well on unseen test data.

\paragraph{Prediction update}
We will start by embedding the \emph{prediction update} in a \gls{rkhs}. In Hilbert space, the mapping of the \textit{a posteriori} mean embedding $\vec{\mu}_{\varphi_t}$ 
of time point $t$ to the prior mean embedding of the next state $\vec{\mu}_{\varphi_{t+1}}^-$\footnote{The hyphen denotes that the mean is an \textit{a priori} belief.} 
is given by a conditional operator $\mathcal{C}_{\varphi'|\varphi}$.  
As our model is unknown, $\mathcal{C}_{\varphi'|\varphi}$ needs to be computed using a sample-based estimator. This can be achieved as
\begin{align}
\mathcal{C}_{\varphi'|\varphi} = \vec{\Upsilon}_{x'}\, (\vec{K}_{xx} + \lambda_T \vec{I}_m)^{-1}\, \vec{\Upsilon}_x^T,
\end{align}
where $\vec{K}_{xx} = \vec{\Upsilon}_x^T\, \vec{\Upsilon}_x$. 
The matrix $\vec{\Upsilon}_x = [\varphi(x_1),\, \dots,\, \varphi(x_m-1)]$ contains the feature mappings of all states 
and $\vec{\Upsilon}_{x'} = [\varphi(x_2),\, \dots,\, \varphi(x_m)]$ the mappings of all subsequent states. 
Therefore, the conditional operator is able to map the mean embeddings of the current state to the embedding of the next state. 
The parameter $\lambda_T$ is used to regularize the observation model (Gram matrix).
The prediction update equations of the traditional \gls{kf} can then be reformulated in a \gls{rkhs} as
\begin{align}
\label{eq:pred_update}
\vec{\mu}_{\varphi_{t+1}}^- 
= \mathcal{C}_{\varphi'|\varphi}\, \vec{\mu}_{\varphi_{t}}
&&
\vec{\Sigma}_{\varphi_{t+1}}^- 
= \mathcal{C}_{\varphi'|\varphi}\, \vec{\Sigma}_{\varphi_{t}}\, \mathcal{C}_{\varphi'|\varphi}^T + \vec{\Upsilon}_{x'}\, \vec{V}\, \vec{\Upsilon}_{x'}^T
\end{align}	
where $\vec{\Upsilon}_{x'}\, \vec{V}\, \vec{\Upsilon}_{x'}^T$ denotes the covariance of the zero-mean Gaussian noise of the transition model which is 
also learned from the training data set. 

\paragraph{Innovation update}
	
For the \emph{innovation update}, we define an observation operator $\mathcal{C}_{\phi|\varphi}$ with 
$\phi(y_t) = \mathcal{C}_{\phi|\varphi} \varphi(x_t) + \nu$, where $\nu$ is zero-mean Gaussian noise. 
Thus, the operator maps the state embedding to the observation embedding and therefore, represents the  \gls{fgkkf} 
equivalent of the observation matrix in the original \gls{kf} formulation. 
The observation operator is estimated using the training data with 
$\mathcal{C}_{\phi|\varphi} = \vec{\Phi}_{y}\, (\vec{K}_{xx} + \lambda_O \vec{I_m})^{-1}\, \vec{\Upsilon}_x^T$, where
 $\vec{\Phi}_{y} = [\phi(y_1),\, \dots,\, \phi(y_m)]$, and $\lambda_O$ is again a regularization parameter. 
In the original \gls{kf} equations, the \textit{a priori} mean and covariances are used in the innovation update 
together with the current observation $y_t$ to compute the Kalman gain, which represents the relative importance of the error with 
respect to the prior belief state estimation. Using the Kalman gain, we finally arrive at the \textit{a posteriori} belief over the state. 
The Hilbert space equivalent of the innovation update equations are 
	\begin{align}
	\vec{\mu}_{\varphi_t} 
	= \vec{\mu}_{\varphi_t}^- + \mathcal{Q}_t\, (\phi(y_t) - \mathcal{C}_{\phi|\varphi}\, \vec{\mu}_{\varphi_t}^-)
	&&
	\vec{\Sigma}_{\varphi_t} 
	= \vec{\Sigma}_{\varphi_t}^- - \mathcal{Q}_t\, \mathcal{C}_{\phi|\varphi}\, \vec{\Sigma}_{\varphi_t}^-
	\end{align}
	where the Kalman gain matrix is computed by
	\begin{align}
	\mathcal{Q}_t = \vec{\Sigma}_{\varphi_t}^-\, \mathcal{C}_{\phi|\varphi}^T\, (\mathcal{C}_{\phi|\varphi}\, \vec{\Sigma}_{\varphi_{t}}^-\, \mathcal{C}_{\phi|\varphi}^T + \kappa \vec{I}_m)^{-1}.
	\end{align}
    
The zero-mean Gaussian noise of the observation model is estimated as $\kappa \vec{I}_m$.

\paragraph{State reconstruction}
All computed means and covariances lie in the Hilbert space. Thus, an additional step is needed to map the embeddings back to the original state space. 
For this \emph{reconstruction of the state distribution} another conditional operator $\mathcal{C}_{X|\varphi}$ is used. 
Similarly to the already defined conditional operators, it is computed using a sample-based estimator resulting in
	\begin{align}
	\mathcal{C}_{X|\varphi} = \vec{X}\, (\vec{K}_{xx} + \lambda_O \vec{I}_m)^{-1}\, \vec{\Upsilon}_x^T
	\end{align}
	with the hidden state matrix $\vec{X} = [x_1,\, \dots,\, x_m]$. The reconstruction is now conducted by applying the conditional operator to the mean and covariance embeddings, yielding
	\begin{align}
	\vec{\mu}_{x_{t}} = \mathcal{C}_{X|\varphi}\, \vec{\mu}_{\varphi_{t}}
	&&
	\vec{\Sigma}_{x_{t}} = \mathcal{C}_{X|\varphi}\, \vec{\Sigma}_{\varphi_{t}}\, \mathcal{C}_{X|\varphi}^T.
	\end{align}
	\subsection{Finite-sample RKHS Embedding}
	The \gls{fgkkf} embeds the state belief in a potentially infinite-dimensional Hilbert space using a non-linear feature map. 
	In this high-dimensional space, non-linear inference can be performed by linear matrix operations as shown above.
	
        As our models are unknown, $\mu_{\varphi_t}$ and $\Sigma_{\varphi_t}$ cannot be computed directly and need to be estimated. 
        The estimation is done by representing the mean embedding at time point $t$ only by a finite vector 
        $m_t \in \mathbb{R}^{m\times1}$ and the covariance by a finite-dimensional matrix $\vec{S}_t \in \mathbb{R}^{m \times m}$ through
	\begin{align}
	\vec{\mu}_{\varphi_{t}} = \vec{\Upsilon}_{x'}\, m_t
	&&
	\vec{\Sigma}_{\varphi_{t}} = \vec{\Upsilon}_{x'}\, \vec{S}_t\, \vec{\Upsilon}_{x'}^T.
	\end{align}	
    
	When inserted into the Equations \ref{eq:pred_update}, we receive a \emph{finite-sample prediction update} 
	formulation where the finite-dimensional \textit{a priori} mean embedding is computed as
	\begin{align}
	m_{t+1}^- &= \vec{T}\, m_{t}.
	\end{align}	
    
	The transition matrix $\vec{T} = (\vec{K}_{xx} + \lambda_T \vec{I}_m)^{-1}\, \vec{K}_{xx'}$ 
	is the finite-dimensional equivalent to the conditional operator $\mathcal{C}_{\varphi'|\varphi}$ 
	and forms the learned model of the underlying system's dynamics. For the covariance embedding estimation we receive
	\begin{align}
	\vec{\Sigma}_{\varphi_{t+1}}^- &= \mathcal{C}_{\varphi'|\varphi}\, \vec{\Sigma}_{\varphi_{t}}\, \mathcal{C}_{\varphi'|\varphi}^T + \vec{\Upsilon}_{x'}\, \vec{V}\, \vec{\Upsilon}_{x'}^T\\
	\vec{S}_{t+1}^- &= \vec{T}\,  \vec{S}_t\, \vec{T}^T + \vec{V}.
	\end{align}	
    
	As a next step, the equations for a \emph{finite-sample innovation update} are introduced. 
	The finite-dimensional Kalman gain matrix $\mathcal{Q}_t \in \mathbb{R}^{m \times m}$ is estimated by
	\begin{align}
	\vec{Q}_t = \vec{S}_t^-\, \vec{O}^T\, (\vec{G}_{yy}\, \vec{O}\, \vec{S}_t^-\, \vec{O}^T + \kappa\, \vec{I}_m)^{-1}
	\end{align}
	where $\vec{G}_{yy} = \vec{\Phi}_y^T\, \vec{\Phi}_y$ is the Gram matrix of the embedded observations. 
	The learned observation model of the underlying system is given by $\vec{G}_{yy}\vec{O}$, where $\vec{O} = (\vec{K}_{xx} + \lambda_O \vec{I}_m)^{-1}\, \vec{K}_{xx'}$. The finite-dimensional Kalman gains can be extracted from the infinite-dimensional ones as $\mathcal{Q}_t = \vec{\Upsilon}_{x'}\, \vec{Q}_t \vec{\Phi}_y^T$ . 
	Using $\vec{Q}_t$, the finite-dimensional \textit{a posteriori} mean embedding is derived as
	\begin{align}
	m_t &= m_t^- + \vec{Q}_t\, (k_{:y_t} - \vec{G}_{yy}\, \vec{O}\, m_t^-).
	\end{align}	
    
	The observations are represented by the kernel vector $k_{:y_t} = [k(y_1,y_t),\, \dots,\, k(y_m,y_t)]$. The finite-dimensional \textit{a posteriori} covariance embedding is then calculated by
	\begin{align}
	\mathcal{S}_t &= \mathcal{S}_t^- - \vec{Q}_t\, \vec{G}_{yy}\, \vec{O}\, \mathcal{S}_t^-.
	\end{align}	
	As for the infinite-dimensional case, the \emph{reconstruction of the state distribution} is needed to map the mean and covariance embeddings back to the original space. For the mean and covariance, the derivations are 
	\begin{align}
	\vec{\mu}_{x_{t}} 
	= \vec{X}\, \vec{O}\, m_t
  	&&
    \vec{\Sigma}_{x_{t}} = \vec{X}\, \vec{O}\, \vec{S}_t\, \vec{O}^T\, \vec{X}^T.
	\end{align}	
	
	\subsection{Sub-space \ac{fgkkf}}
	\label{sec:sub-space_fgkkf}
The \gls{kkr} possesses almost cubic computational complexity for the number of training samples due to the inversion of the Gram matrix $\vec{K}_{xx} \in \mathbb{R}^{m \times m}$ \cite{KKR}. 
Thus, for large training sets the calculations become practically intractable. 
However, a sub-space \gls{fgkkf} variant can be defined that allows to work with large data sets. 
The core idea is to represent the mean embedding only by a subset of the training samples, while still all samples are used to learn the model. 
To achieve this, a sub-space feature mapping $\vec{\Gamma}_x = [\varphi(x_1),\, \dots,\, \varphi(x_{n-1})]$ which contains the mappings of only $n \ll m$ training samples is defined, such that $\vec{\Gamma}_x \subset \vec{\Upsilon}_x$. 
The Gram matrix is then calculated as $\vec{\overline{K}_{xx}} = \vec{\Upsilon}_x^T\, \vec{\Gamma}_x$ with dimensions $\vec{\overline{K}_{xx}} \in \mathbb{R}^{m \times n}$  leading to new conditional embedding operators for the model learning which are called sub-space conditional embedding operators, introduced in \cite{KKR}.
	
The formulations for the \emph{prediction update} of the sub-space \gls{fgkkf} stay the same as for the full-space \gls{fgkkf} with
\begin{align}
\vec{\mu_{\varphi_{t+1}}^-} 
= \mathcal{C}^{S}_{\varphi'|\varphi}\, \vec{\mu_{\varphi_{t}}}
&&
\vec{\Sigma}_{\varphi_{t+1}}^- 
= \mathcal{C}^{S}_{\varphi'|\varphi}\, \vec{\Sigma}_{\varphi_{t}}\, \mathcal{C}^{ST}_{\varphi'|\varphi} + \vec{\Upsilon}_{x'}\, \vec{V}\, \vec{\Upsilon}_{x'}^T
\end{align}	
but the conditional operator is now given by $\mathcal{C}^{S}_{\varphi'|\varphi} = \vec{\Upsilon}_{x'}\, \vec{\overline{K}}_{xx}\, \vec{L}_T^S\, \vec{\Gamma}_x^T$, where $\vec{L}_T^S = (\vec{\overline{K}}_{xx}^T\, \vec{\overline{K}}_{xx}  + \lambda_T \vec{I}_n)^{-1}$. 
\\
For the \emph{innovation update} also the formulation of the the sub-space \gls{fgkkf} stay the same as for the full-space \gls{fgkkf} with the conditional operator mentioned before:
\begin{align}
\vec{\mu}_{\varphi_t} 
= \vec{\mu}_{\varphi_t}^- + \mathcal{Q}^S_t\, (\phi(y_t) - \mathcal{C}^{S}_{\phi|\varphi}\, \vec{\mu}_{\varphi_t}^-)
&&
\vec{\Sigma}_{\varphi_t} 
= \vec{\Sigma}_{\varphi_t}^- - \mathcal{Q}^S_t\, \mathcal{C}^{S}_{\phi|\varphi}\, \vec{\Sigma}_{\varphi_t}^-.
\end{align}
\section{Putting the FKKF to Work}
\label{sec:gkkf_algo}
For the traffic prediction experiments described in Section~\ref{sec:evaluation}, we use the sub-space \gls{fgkkf} in order to make the computations tractable even when using a high number of training samples. This section presents our realization of the sub-space \gls{fgkkf}, shown as pseudocode in Algorithm~\ref{alg:subspace_gkkf}, and its implementation on state of the art data center switching devices.

    \subsection{Preprocess Data}
\label{sec:gkkf_algo_preprocess_data}
The main idea of preprocessing the data, is that the time series are seen as a time-discrete signal sampled using a sampling interval $T_{S}$. 
Instead of learning with the signal in the time domain we transform it to the frequency domain by making use of the \ac{ft}. 
An advantage of conducting the calculations in the frequency domain is that since the \ac{ft} is built over a part of the signal, much more information about the flow structure is contained in every data point which makes it easier to observe the current condition of the flow. 
Furthermore, since $T_{C} > T_{S}$ the data set will contain less data points in the frequency than in the time domain. 
This is especially useful when the training set consists of very long flows that would otherwise result in a large number of data points. 
The selection of an appropriate value for $T_{C}$ that leads to an overlap of the chunks also reduces artifacts in the frequency domain.

In the end, the resulting predictions are brought back to the time domain so that a \ac{te} system would receive them as kbit values which it could then use to compute optimal paths for the flows and reroute them accordingly. For representing the signal in the frequency domain, it first has to be split into smaller overlapping chunks of a certain length $w$. The starting points of the single chunks at which they are cut out of the original signal are defined by the sampling interval $T_{C}$. Consequently, the parameters $T_{S}$, $T_{C}$ and $w$ decide  how many chunks the original signal is split into, and by how much the chunks overlap. Since the signal is discrete in time the \ac{dft} needs to be used for the transformation and by splitting the signal in chunks, basically a discrete-time \ac{stft} is performed, whereas a \ac{fft} algorithm is used for making the \ac{ft} computations faster (cf.~\simult).
    
    \subsection{Learning Phase}
    \label{sec:learningPhase}
When using the sub-space \gls{fgkkf} for state estimation, the first step is to choose different settings for the execution of the algorithm. 
This includes choosing a training set $data_{train}$ and setting the values of multiple hyperparameters that are needed for the calculations and whose values are found through optimization in the learning phase.
The hyperparameters $\lambda_T$ and $\lambda_O$ (see Eqs. 1 and 2 respectively) are used for regularizing the Gram matrix products before their inversion; the former parameter is used for the transition model during the prediction update step and the latter for the observation model during the innovation update step. 
The bandwidth scaling factors $state_{bw}$ and $obs_{bw}$ decide which values the bandwidths of the kernel functions --- 
used for the state and observation embeddings --- 
are multiplied with. 
The bandwidths are found by choosing a subset of training samples and calculating the median of the squared distances between them. 
The last hyperparameter is the observation noise covariance $\kappa$ (see Eq. 4). 
Besides forming a state representation out of observations, and thus allowing the \ac{fgkkf} to process data with a very high dimensionality, the \textsc{preprocess-data} function includes conducting a \gls{pca} on the training set. The \gls{pca}, which is conducted on a dataset after bringing it to the frequency domain and after standardizing it so that the set possesses zero mean and unit variance,  reduces the dimensions needed for a single observation to lower their complexity and save computation time. 
The \ac{pca} uses an orthogonal transformation to compute \acp{pc} corresponding to the dimensions of the dataset that are ordered in a descending order, meaning the first \ac{pc} explains more variance of the dataset than any other \ac{pc}. 
The assumption is that some frequencies might account for such a small amount of variance that their omission distorts the characteristics of the flow only insignificantly. 
It is important to mention that the \ac{pca} must always be conducted on the training set. The test set which is usually not fully available  
also needs to be standardized and its dimensionality reduced. However, for the transformation, the \acp{pc} computed from the training set must be used.

\subsection{\ac{rkhs} Embedding of the Kalman Filter}
 \label{sec:rkhskf}

In the \textsc{learn} function the Gram matrices of the state embeddings denoted ${\overline{K}}_{xx}$ and ${\overline{K}}_{xx'}$ (see Section~\ref{sec:sub-space_fgkkf}) 
are formed together with the observation Gram matrix ${G}_{yy}$. The matrices are then used to estimate the sub-space transition model matrix ${T}^S$ and the sub-space observation model matrix ${GO}^S$. Moreover, the state matrix ${X}$ needed for the projection back to the state space is formed and the initial values for the \textit{a priori} sub-space mean embedding $\vec{n}_1^-$ and covariance embedding $\vec{P}_1^-$ are set.

\begin{table}[ht]
\small
\centering
\begin{tabular}{|c|l|}
\hline
\rule{0pt}{14pt}
\bf{Parameter} & \bf{Description}\\
\hline
\rule{0pt}{14pt}
$data_{train}$ & Training set\\
\hline
\rule{0pt}{14pt}
$\lambda_T$ & Regulation parameter for the transition model\\
\hline
\rule{0pt}{14pt}
$\lambda_O$ & Regulation parameter for the observation model\\
\hline
\rule{0pt}{14pt}
$state_{bw}$ & Scaling factors for the state embeddings\\
\hline
\rule{0pt}{14pt}
$obs_{bw}$ & Scaling factors for the observation embeddings \\
\hline
\rule{0pt}{14pt}
$\kappa$ & Observation noise covariance\\
\hline
\rule{0pt}{14pt}
${\overline{K}}_{xx}$, ${\overline{K}}_{xx'}$ & Gram matrices of the state embeddings\\
\hline
\rule{0pt}{14pt}
${G}_{yy}$ & Observation Gram matrix\\
\hline
\rule{0pt}{14pt}
${T}^S$ & Sub-space transition model matrix\\
\hline
\rule{0pt}{14pt}
${GO}^S$ & Sub-space observation model matrix\\
\hline
\rule{0pt}{14pt}
${X}$ & State matrix\\
\hline
\rule{0pt}{14pt}
${P}_{t}$ & Covariance embedding\\
\hline
\rule{0pt}{14pt}
${Q}^{S}_{t}$ & Sub-space Kalman gain matrix\\
\hline
\rule{0pt}{14pt}
${n}_{t}$ & Sub-space mean embedding\\
\hline
\rule{0pt}{14pt}
$\mu_{x_t}$ & Mean prediction\\
\hline
\rule{0pt}{14pt}
$\Sigma_{x_t}$ & Covariance prediction\\
\hline
\end{tabular}
\caption{Parameters of the FKKF Algorithm}
\label{table:paraFGKKF}
\end{table}

    \subsection{Prediction Phase}
The \gls{fgkkf} algorithm can either be used online or offline.  When used online, single observations would be omitted from the underlying system and directly used in the innovation update step. During the traffic prediction experiments described in Section~\ref{sec:evaluation} the algorithm was used offline. 
In the equations for the prediction and innovation updates shown in Section~\ref{sec:gkkf}, the Kalman gain matrices and the covariances were calculated together with the mean embeddings. However, none of the components depend on the current observation $y_t$ and therefore, they can be calculated before receiving any observation. Thus, the calculation can be done in the function \textsc{project} before testing the model which reduces the computation time needed for the \gls{fgkkf} algorithm  (cf.~\simult). 
In the offline mode, the projection of ${Q}_t^S$, $P_t$ and $P_{t+1}^-$ allows to save computation time because when testing the model simultaneously with multiple test sets the components only have to be calculated once. 
The complexity of the prediction procedure has the complexity of a matrix-matrix multiplication with $O(n^3)$ where $n$ here depends on the number of samples combined with the diversity of the flow structure, yet is \emph{independent of the number of flows and the number of network elements}. 
There are also algorithms available with a lower complexity but the used algorithm is a block matrix algorithm where the blocks are small enough to fit into local memory which reduces the shifts into and out of memory. 

In the last function the model is tested with given data, where the \textsc{project} function is called once before the prediction phase starts.
The only step left in the \textsc{predict} function, for the innovation and prediction update, is the calculation of the \textit{a posteriori} and the \textit{a priori} sub-space mean embeddings $n_t$ and $n_{t+1}^-$, respectively. 
After the prediction, the mean and covariance embeddings are projected back into the state space. 
The prediction and innovation update steps can be executed alternately. However, if the observations arrive irregularly only every $p$-th time step, the algorithm will solely execute the prediction update step and thus perform a $p$-step prediction of the system state.

\begin{algorithm}[!ht]
	\caption{Core \ac{fgkkf} algorithm}
	\label{alg:subspace_gkkf}
\begin{algorithmic}
 	\Function{preprocess-data}{$data$}
 	\State $stateWindow$ $\leftarrow$ \textsc{form-state-windows}($data$) 	    	\State $data$ $\leftarrow$ \textsc{FFT}($stateWindow$)
    \State \Return $data$
 	\EndFunction
    \vspace{1mm}
    	\Function{learn}{$data_{train}$}
    \While{$data$ in $data_{train}$}
	\State $data$ $\leftarrow$ \textsc{preprocess-data}($data$)
    \State $dimensionReducedData$ $\leftarrow$ \textsc{pca}($data$)
    \State $\langle \lambda_T$, $\lambda_O$, $state_{bw}$, $obs_{bw}$, $\kappa\rangle$ $\leftarrow$\textsc{optimize-hyperparams}
    ($dimensionReducedData$)
    \State $\langle \vec{T}^S$, $\vec{GO}^S$, $\vec{X}$, $\vec{n}_{1}^-$, $\vec{P}_{1}^-\rangle$ $\leftarrow$ \textsc{estimate-model}($\vec{\overline{K}}_{xx}$, 
    $\vec{\overline{K}}_{xx^{'}}$, $\vec{G}_{yy}$, $dimensionReducedData$)
    \EndWhile
    \State \Return $\langle \lambda_T$, $\lambda_O$, $state_{bw}$, $obs_{bw}$, $\kappa$, $\vec{T}^S$, $\vec{GO}^S$, $\vec{X}$, $\vec{n}_{1}^-$, $\vec{P}_{1}^-$, 
$featureSet\rangle$
	\EndFunction
    \vspace{1mm}
   	\Function{project}{$learnedPara$}
   	\State ${Q}_t^S$ $\leftarrow$ \textsc{kalman-gain}($learnedPara$)
	\State ${P}_t$ $\leftarrow$ \textsc{a-posteriori-covariance-embed}($learnedPara$)
	\State ${P}_{t+1}^-$ $\leftarrow$ \textsc{a-priori-covariance-embed}($learnedPara$)
    \State \Return $\langle {Q}_t^S$, ${P}_t$, ${P}_{t+1}^-\rangle$
	\EndFunction
    \vspace{1mm}
   	\Function{predict}{$data$, ${Q}_t^S$, ${P}_t$, ${P}_{t+1}^-$}
   	\State $data$ $\leftarrow$ \textsc{preprocess-data}($data$)
	\State Innovation update:
	\State 
  $\vec{n}_t$$\leftarrow$\textsc{a-posteriori-mean-embedding}($data$, ${Q}_t^S$, ${P}_t$, ${P}_{t+1}^-$)
	\State Prediction update:
	\State 
    $\vec{n}_{t+1}^-$$\leftarrow$\textsc{a-priori-mean-embedding}   ($data$, ${Q}_t^S$, ${P}_t$, ${P}_{t+1}^-$)
	\State Project into state space:
	\State 
  $\vec{\mu}_{x_t}$$\leftarrow$\textsc{mean-predict}($data$, $n_t$, $n_{t+1}^-$, ${Q}_t^S$, ${P}_t$, ${P}_{t+1}^-$)
    \State 
    $\vec{\Sigma}_{x_t}$$\leftarrow$\textsc{covariance-predict}($data$, $\vec{n}_t$, $\vec{n}_{t+1}^-$, ${Q}_t^S$, ${P}_t$, ${P}_{t+1}^-$)
    \State \Return $\langle \vec{\mu}_{x_t}$, $\vec{\Sigma}_{x_t}\rangle$
	\EndFunction
    \vspace{1mm}
   	\Procedure{test-model}{$data_{test}$}
   	\State $\langle\vec{Q}_t^S$, $\vec{P}_t$, $\vec{P}_{t+1}^-\rangle$ $\leftarrow$ \textsc{project}($ \lambda_T$, $\lambda_O$, $state_{bw}$, $obs_{bw}$, $\kappa$, $\vec{T}^S$, 
$\vec{GO}^S$, $\vec{X}$, $n_{1}^-$, $P_{1}^-$, $featureSet$)
    \While{$data$ in $data_{test}$}
    \State $\langle \vec{\mu}_{x_t}$, $\vec{\Sigma}_{x_t}\rangle$ $\leftarrow$ \textsc{predict}($data$, ${Q}_t^S$, $\vec{P}_t$, $\vec{P}_{t+1}^-$)
    \State \textsc{check-results}~($\vec{\mu}_{x_t}$, $\vec{\Sigma}_{x_t}$, $data+1$)
    \EndWhile
	\EndProcedure
\end{algorithmic}
\end{algorithm}
\subsection{System Design}
\label{sec:sys_design}

\begin{figure*}
\centering
\includegraphics[width=180mm]{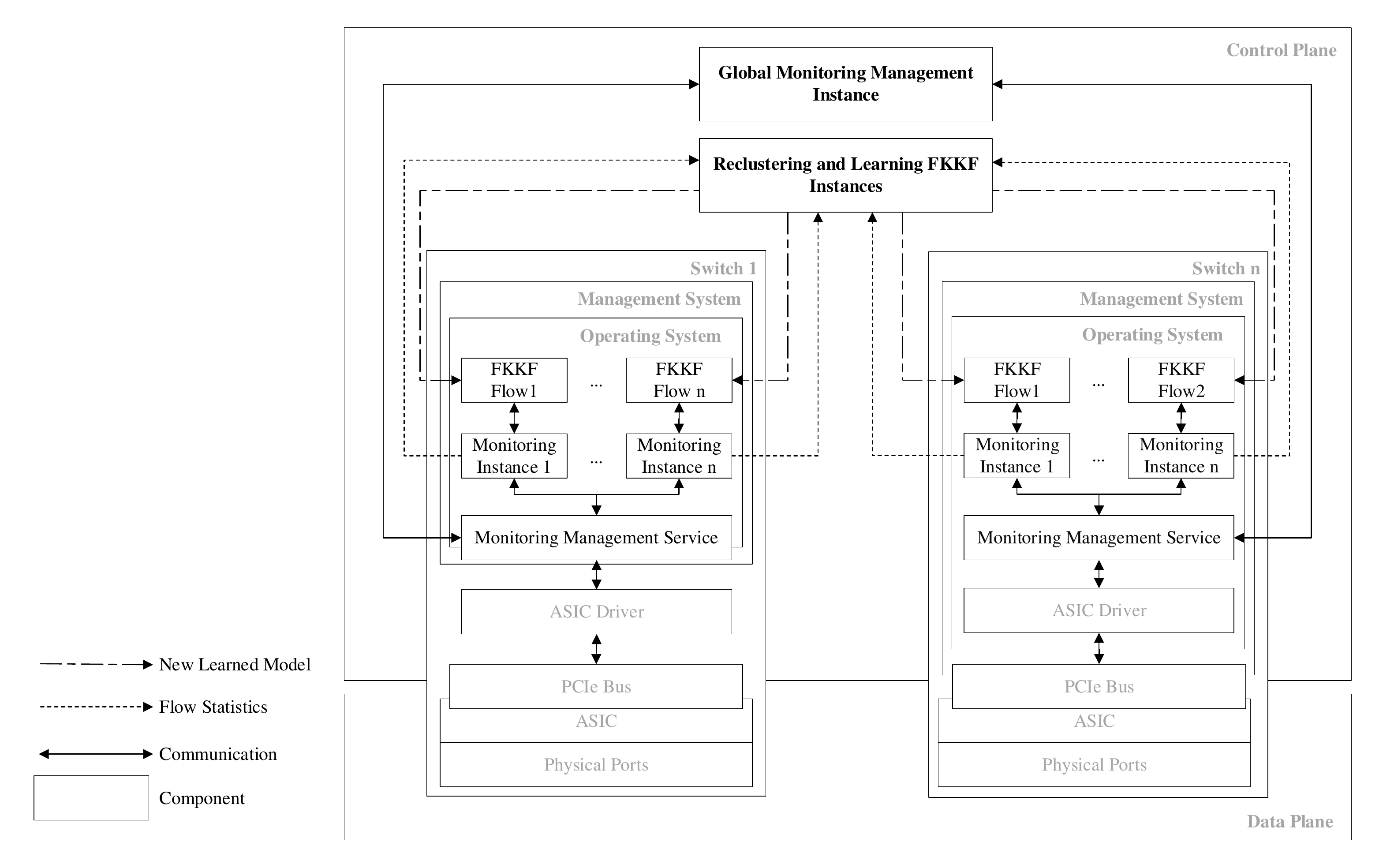}
\caption{
Overview of the \ac{fgkkf} system. The control plane summarizes the management systems of different switches, while the \ac{asic} is responsible for fast packet forwarding  as part of the data plane. A PCIe bus typically connects the \ac{asic} and the management system where usually a Linux based OS is running with the communication interface (an \ac{asic} driver) installed. This \ac{asic} driver updates the \ac{tcam} and fetches the statistics from the \ac{asic}.  The monitoring system is implemented above the \ac{asic} driver with its components (monitoring management service, monitoring instances) which  communicate with a global monitoring management instance. The prediction algorithm runs on top of a monitoring instance and predicts the future traffic of individual flows. The monitoring instance sends a copy of the current statistics to a centralized reclustering and learning \ac{fgkkf} instance. This instance starts a new clustering and learning process if needed and submits the new model to the running \ac{fgkkf} instance.
}
\label{fig:switch}
\end{figure*}

The requirements for fine grained prediction (cf. \nonlin) 
lead to two immediate consequences when implementing prediction algorithms:
\begin{itemize}
\item The interval for fetching  statistics of individual flows has to be small enough to allow the algorithm to learn the fine grained structure of a flow (in our case 10\msecs).
\item The prediction and further steps have to be ``close'' to the forwarding plane to minimize latency. Predicting more in advance is likely to compromise on accuracy and so its more efficient to set the prediction time optimal with respect to accuracy, yet avoid additional latency between the signal, the prediction and further steps.
\end{itemize}
State of the art data center switching devices, as shown in Figure~\ref{fig:switch}, have two main processing instances -- ($i$) a management system (with a common \acrshort{cpu}), and ($ii$) a forwarding \ac{asic}. While the \ac{asic} is optimized for fast packet forwarding, the management system communicates with other control plane instances (e.g., other switch management systems,  \acl{sdn} controller) and updates the \ac{tcam} where the forwarding rules of the \ac{asic} are maintained. 
Another task of the management system is to fetch statistics or packet samples from the \ac{asic}.
The \ac{asic} provides statistics of all forwarding rules, of individual ports and aggregated statistics of the forwarding system.
Updating the \ac{tcam}, fetching the statistics, or receiving packet samples are done over a \ac{pcie} bus.
On current white box switches, the \emph{management system} typically runs a Linux-based \ac{os}.
This \ac{os} can be open source (i.e. \ac{onl}\footnote{http://opennetlinux.org/}, OpenSwitch \footnote{https://www.openswitch.net/}) or extensible (i.e. Arista EOS\footnote{https://www.arista.com/en/products/eos}). 
To be able to implement the \ac{fgkkf}, a new monitoring system had to be developed because existing approaches like sFlow~\cite{sFlow} and IPFIX~\cite{IPFIX} are designed for fetching the data of interest from a switch and sending them to a centralized instance where they are analyzed.
This process, even if all routes are globally optimized, takes too long because of the short prediction time as  mentioned  above. 
The new monitoring system now offers the possibility for complex processing of the monitored data directly on the switch. In short our monitoring system consists of ($a$) \emph{monitoring instances} for each flow ($b$) one \emph{monitoring management service} on each switch, and ($c$) a single global \emph{monitoring management instance} 
 (see Figure~\ref{fig:switch}).
The monitored data from the \ac{asic} is forwarded over a monitoring management service which aggregates queries from different monitoring instances and sends the required data back to the monitoring instances.
One of the mentioned individual monitoring instances can deliver the data to a \ac{fgkkf} instance to predict the trajectory of a given flow.
The monitoring instances have their own state, which allows them to collect statistics, process them,  and store data.
The global monitoring instance forms a management system which communicates to the monitoring management services on the switches to synchronize and optimize the monitoring utilization.
Note that the statistics from the monitoring instance is not only used for 
online prediction of the flow directly on switches, but also for 
offline classification and relearning of the flow structure (if needed, see~\ref{sec:learningPhase}). 
The prediction algorithm runs on the management system of the switch to receive the prediction as quickly as possible. 
Once the clustering and learning process is completed, the reclustering and learning \ac{fgkkf} instance updates the prediction model (described in~\ref{sec:gkkf_algo}) of the \ac{fgkkf} instance which is running on the switch.
Such a system can now be used for new \ac{te} approaches as shown shortly 
in Section~\ref{sec:evaluation}.
\section{Evaluation}
\label{sec:evaluation}
In this section we assess the performance of our \gls{fgkkf} in traffic flow prediction.

\subsection{Overview and Synopsis} 

To achieve the full potential of the \gls{fgkkf}, and configure its parameters, we have to understand the given data and their nature.
Therefore, the nature of the  data is first analyzed and explained (Section~\ref{sec:exp_data}).  
Then, the data is clustered into  20 different groups by calculating the Euclidean distance between the flows in the frequency domain. 
A subset of 
 the contained flows is either selected for forming training or test sets (cf. \other), and processed by computing the \ac{ft} and using the \ac{pca} to reduce data complexity of the data without loosing too much entropy (Section~\ref{sec:preprcessingData}).
The trade-off between complexity and entropy of the data is optimized to  run the \gls{fgkkf} algorithm on commodity switching hardware, as explained in Section~\ref{sec:sys_design},  withouth increasing the prediction error  (cf. \simult).  
We evaluate both the prediction results 
and the run-time performance (Section~\ref{sec:expResults}).
Note that we predict the exact \emph{trajectory}, which is much more sophisticated  
than probably needed for \ac{te} and congestion control mechanisms, where predicting the highest peak in the next prediction time frame suffices.
By considering the highest peak, the prediction error of the \ac{fgkkf}, as well as of ARIMA and GARCH are calculated (see Table~\ref{table:pred_exp}).
The run time performance measurements on commodity hardware switches shows how accurately the statistics can be polled and how many flows can be predicted on a switch.

\begin{figure}
  \centering
\begin{minipage}[b]{0.8\textwidth}
  ~~~\includegraphics[width=.9\linewidth]{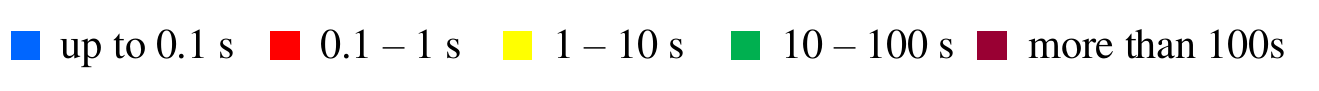}
\end{minipage}
\smallskip
\\
\begin{minipage}[b]{0.35\textwidth}
    \includegraphics{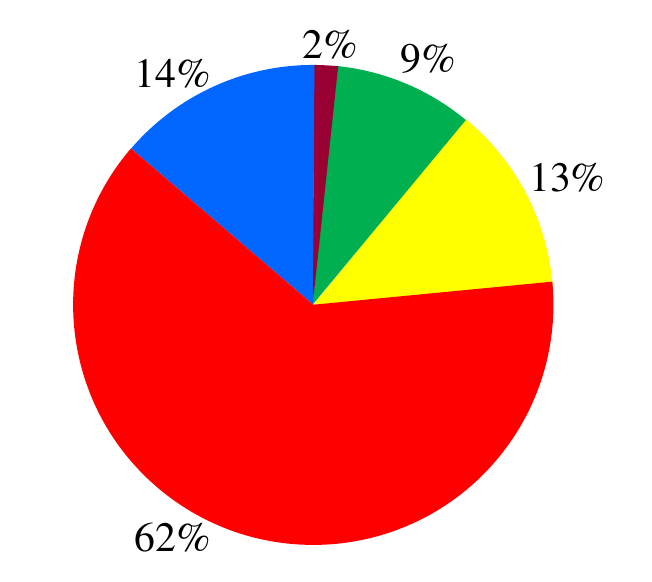}
    \label{fig:http_dur}
  \end{minipage}
\qquad
  \begin{minipage}[b]{0.35\textwidth}
    \includegraphics{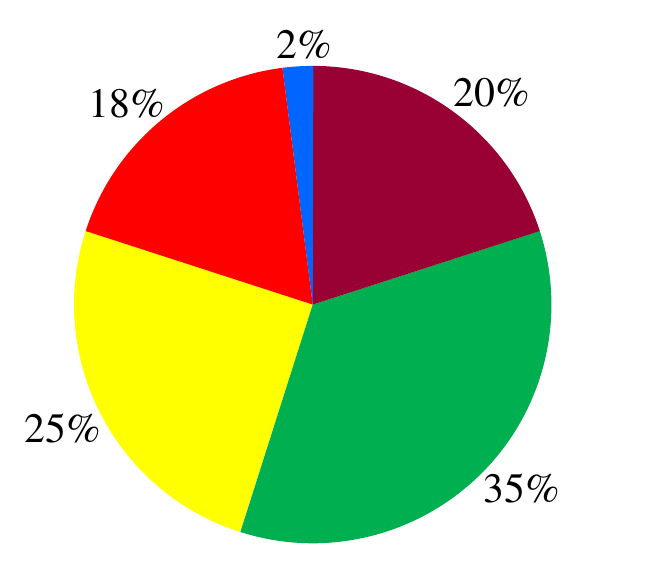}
    \label{fig:http_dur_traffic}
  \end{minipage}
\caption{Characteristics of HTTP flows. Left duration of HTTP flows right Traffic load shares of duration groups}
\label{fig:http_flow_c}
\end{figure}

\subsection{Experimental Data}
\label{sec:exp_data}
The data set used during the experiments was collected as part of a study conducted by Benson et al.\ that aimed at analyzing the characteristics of network traffic in data centers \cite{data_set}. 
During their study, \gls{snmp} data, topology information, and packet traces were collected from university data centers.
Compared to other data sets (e.g.,  from Facebook\footnote{https://research.fb.com/data-sharing-on-traffic-pattern-inside-facebooks-datacenter-network/}) the chosen data set contains information up to the used application layer protocol type. 
We used the packet traces which contain 65 minutes worth of traffic data collected at an edge switch in a university data center consisting of 22 network devices and 500 servers. 

In the first step, the packet traces were analyzed by gathering statistical information about the composition of the traffic data. For 87\% of the traffic transmitted by the observed network switch, TCP was used as the transport layer protocol, and only 13\% was transmitted over UDP. A closer look at the TCP traffic then revealed which protocols were used at the application layer. The \gls{http} traffic makes up the largest share with 75\% of the TCP traffic.
Since the goal is to investigate if the progression of individual flows can be predicted, we focused on traffic using the same application layer protocol because we initially assumed that those flows would already share enough flow characteristics so that knowledge about some of the flows would allow us to predict other unseen ones. The decision was made in favor of \gls{http} traffic because this group contains  65\% of all the collected traffic data. 

However, further analysis of the \gls{http} traffic revealed that individual \gls{http} flows are very heterogeneous. The packet traces contain roughly 130\,000 unique \gls{http} flows, where a flow represents a socket-to-socket connection between two hosts in the data center 
that starts with a TCP handshake and ends with a TCP teardown. 

 \begin{figure}
 	\centering
 	\includegraphics[width=1\textwidth]{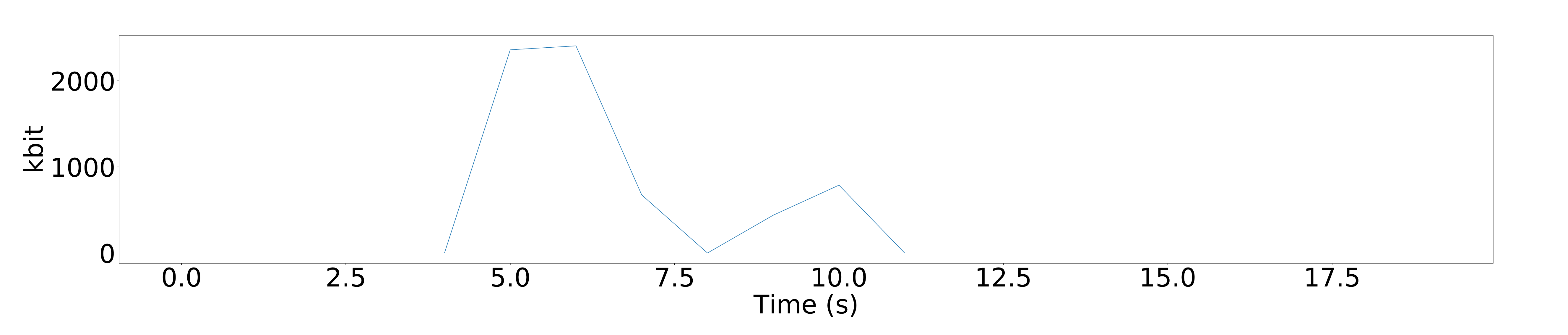}
 	\caption{Sampling interval of 1\secs}
	\label{fig:flow1s}
\end{figure}
\begin{figure}
 	\centering
 	\includegraphics[width=1\textwidth]{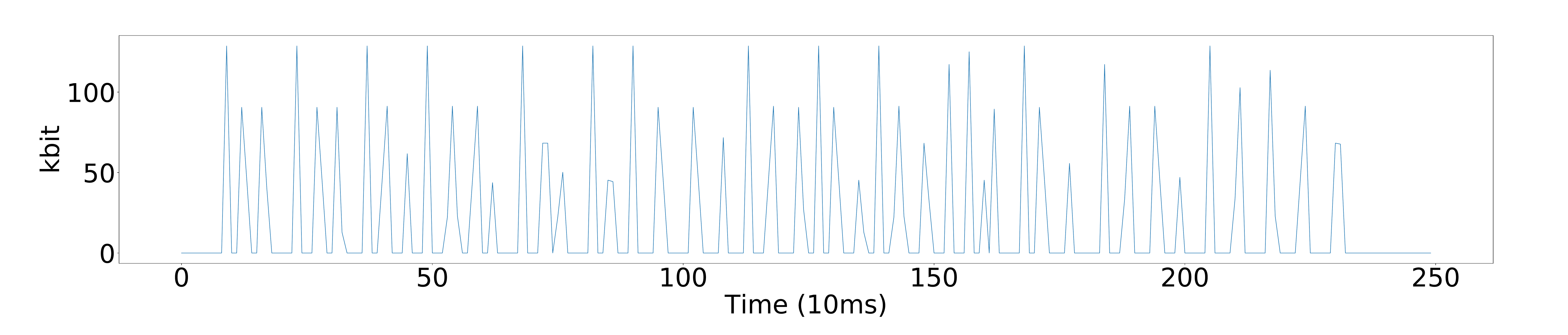}
 	\caption{Fine grained flow structure of the first peak of the figure above with 10ms sampling size}
 	\label{fig:flow001s}
\end{figure}
\label{fig:peak_structure}

As shown in Figure~\ref{fig:http_dur}, 
the durations across \gls{http} flows are highly diverse. Most of the flows (62\%) are very short with a length between 0.1\secs and 1\secs, 
as indicated by the red slice of the pie chart. 
To fully grasp the characteristics of \gls{http} flows it is important to know which share of the traffic load is caused by which \emph{duration group} (group of flows with similar durations).
This information is given in Figure~\ref{fig:http_dur_traffic} and together with Figure~\ref{fig:http_dur} yields  interesting insights. 
The extremely short flows with a duration of up to 0.1\secs cause only 2\% of the traffic and can therefore be neglected. 
The same goes for many flows in the duration group of 0.1\secs to 1\secs, though not for all of them. 
The most important group seems to be the one which contains flows with a rather short length between 1\secs and 100\secs, 
since only 22\% of the flows possess such a length and still they cause 60\% of the traffic. 
These traffic flows could not be captured when aggregating the flows as done in prior work as described in Section~\ref{sec:traffic_pred},  since the flows would only be represented by one or a few data points (cf. \indiv). 
Just the flows longer than 100\secs could be captured sufficiently (see Section~\ref{sec:expResults}).  
As Figure~\ref{fig:http_dur_traffic} indicates, this duration group is responsible for 20\% of the traffic. 
However, 
when examining the progression of high-volume long-lasting flows, a typical flow pattern is observed. 
One example flow is shown in Figure~\ref{fig:flow1s} which clearly reveals the extremely bursty nature of the network traffic.  
Even though an already rather low sampling interval of 1\secs is used,  
no flow structure can be observed since the short but very high peaks are only represented by one or two data points.  
For most of the time, 
the kbit values of the flow are either zero or are insignificant.   
This shows that even the long flows often just consist of short traffic peaks with long low-volume phases between them.  
For a \gls{te} system that tries to predict and reroute traffic in a network this means that it needs to be able to predict the flows on a small time scale because most of the traffic is transmitted in short high-volume traffic peaks. 
This also implies that the traffic data must be represented using a small sampling interval. Figure~\ref{fig:flow001s} shows the first peak of the same flow (Figure~\ref{fig:flow1s}), however this time, a sampling interval of 0.01\secs was used and now much more information about the peak structure can be obtained. 

As we can see, 
a peak itself consists of shorter peaks which  we henceforth refer to as \emph{(peak) impulses}. 
At the lowest possible sampling interval these impulses would represent single packets of the flows. 
In the vast majority of flows, 
the peaks consist of (1) a rising phase in which the peak impulses gradually increase up to a certain maximum kbit value and (2) a peak body where the traffic load remains rather constant. 
In the context of network congestion the rising phase is the most interesting part of the flow because here the traffic pattern of the flow is changing rapidly which causes  congestion. 
Thus, the goal 
should be a successful prediction of the course of the peak rise given only a small part of it. 

Therefore, we next investigate 603 single flows that transmit around 46\% of all the \gls{http} traffic.  

\subsection{Preprocessing the Data}
\label{sec:preprcessingData}
\begin{figure}
 	\centering
 	\includegraphics[width=1\textwidth]{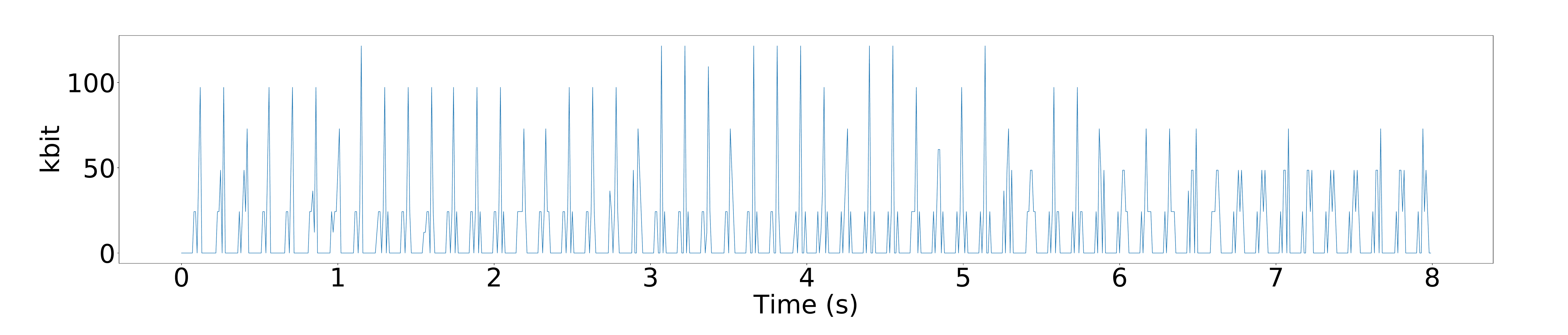}
 \caption{Traffic flows of recurring client-server communication}
 \label{fig:rec_con2}
 \end{figure}

As shown in Figure~\ref{fig:rec_con2}, the flow structure is extremely bursty when observed at a small sampling interval. 
As a consequence, the times series of kbit values produced from the flow packet traces contain many zero-value phases and generally high gradients. 
This makes the course of the time series difficult to learn for any model and is the reason why the approaches described in Section~\ref{sec:traffic_pred} aggregate the flows on a temporal scale and either is not applicable for linear time series with variable variance (ARIMA) and predict zeros or very low values, or start to oscillate arbitrarily (GARCH) because it can not predict the complex hidden state of the system.
However, as we will see in Section~\ref{sec:expResults} this would prevent a \ac{te} system from predicting short traffic peaks. 

  \begin{figure}
 	\centering
 	\includegraphics[width=1\textwidth]{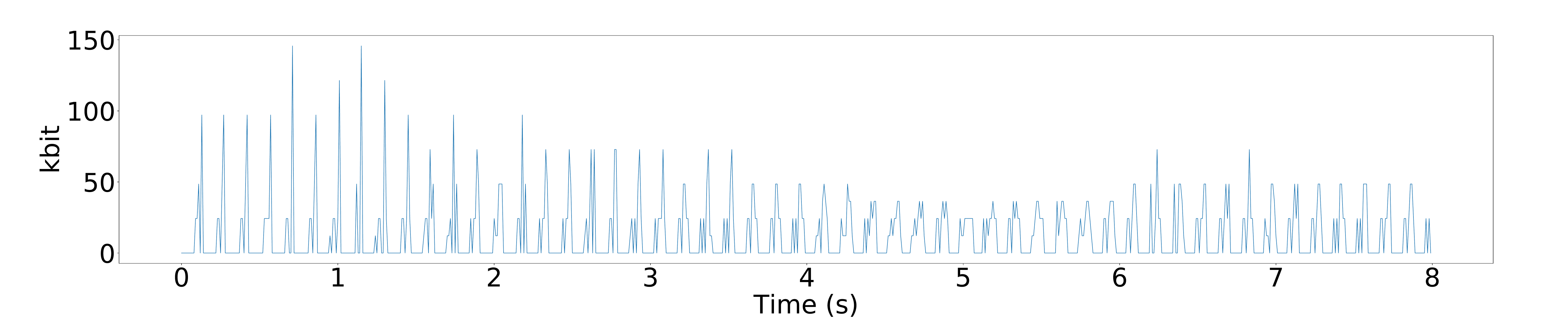}
  \end{figure}

\begin{figure}
 	\centering
 	\includegraphics[width=18cm]{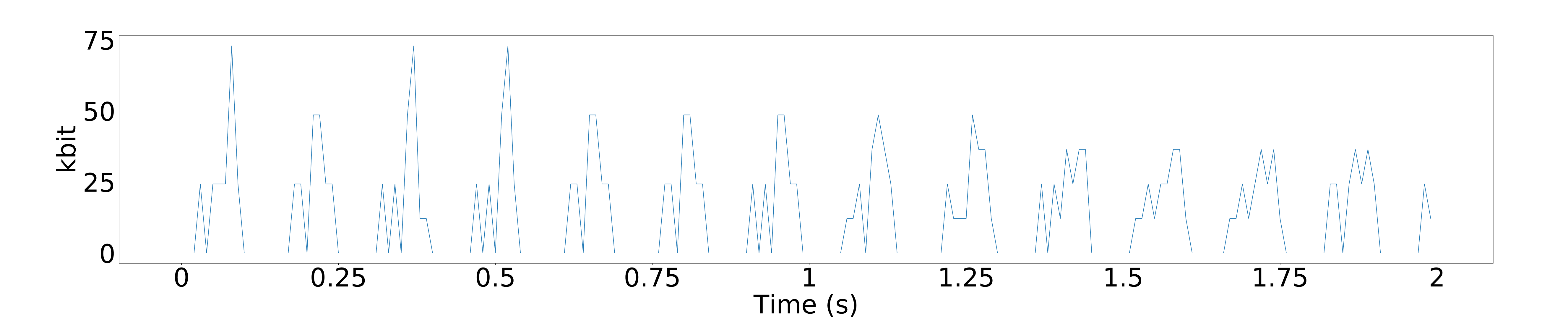}
 	\caption{Signal chunk in the time domain}
 	\label{fig:comp_tf_a}
 \end{figure}

\begin{figure}
 	\centering
 	\includegraphics[width=18cm]{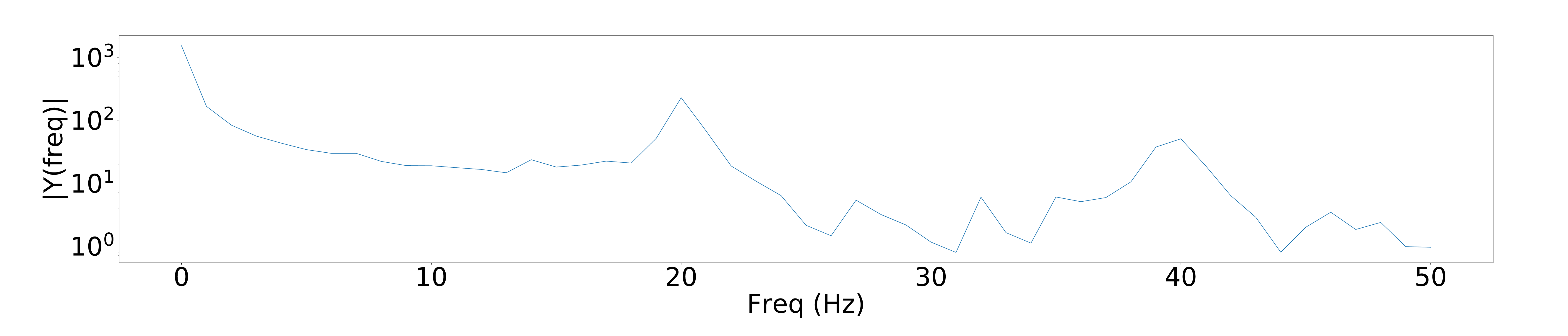}
 	\caption{Signal chunk in the frequency domain}
 	\label{fig:comp_tf_b}
 \end{figure}

\begin{figure}
 	\centering
 	\includegraphics[width=18cm]{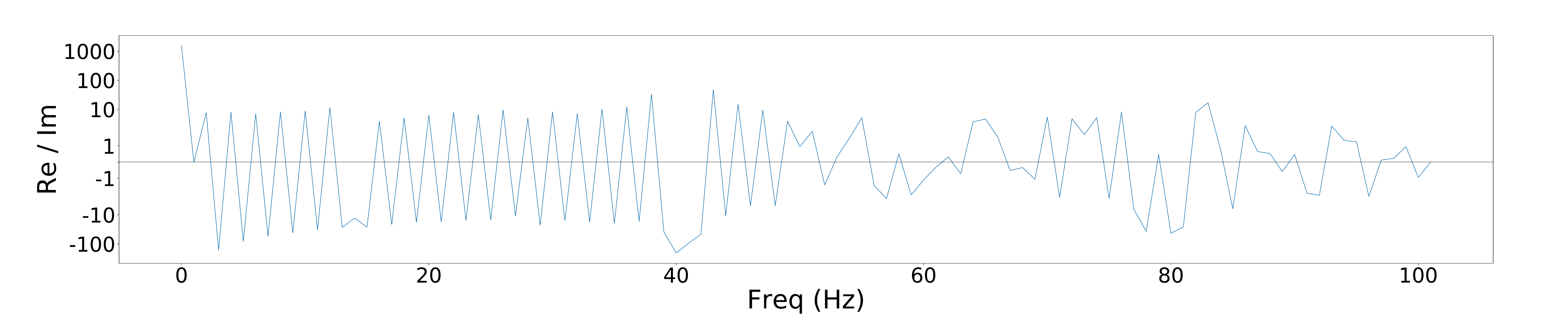}
 	\caption{Signal chunk in the frequency domain with split complex numbers}
 	\label{fig:comp_tf_c}
 \end{figure}

By learning the traffic signal in the frequency instead of the time domain the data sets that the \ac{kkf} has to deal with become less bursty (see Section~\ref{sec:gkkf_algo_preprocess_data}). Figure~\ref{fig:comp_tf} compares the courses of different data series. In Figure~\ref{fig:comp_tf_a} a 2\secs\ long flow chunk is shown. When computing the \ac{ft} of the chunk, a data series as shown in Figure~\ref{fig:comp_tf_b} is produced. The data series consists of complex numbers that describe the amplitudes and phases of simple sine waves that can be combined to construct the original signal. In the plot the amplitudes of the needed sine waves are shown. Furthermore, the plot reveals that the frequency values get folded  at the Nyquist frequency $f_{N}$ which is half of the sampling frequency, $f_{N} = 0.5\, f_{S} = 0.5\, \frac{1}{T_{S}}$. For reconstructing the time signal from sine waves we therefore only need to save the first $N$ values of the \ac{ft} results, with $N = f_{N} + 1$. When transforming our results back to the time domain, the rest of the values can be reconstructed by inverting the imaginary parts of the saved complex numbers. However, in order to use the \ac{ft} results with the \ac{fgkkf}, the complex numbers have to be split into their real and imaginary parts. The resulting data series consisting alternately of real and imaginary values is presented in Figure~\ref{fig:comp_tf_c} which clearly shows that the data series is considerably less bursty than the original time series.

The influence of the transition to the frequency domain on the employed data set should now be illustrated using an example. We assume that the training set that the \ac{kkf} should learn from consists of only one short high-volume flow with a duration of 10\secs. When using a sampling interval $T_{S} = 0.01$ our training set $data_{train}$ contains a time series of kbit values observed every 0.01\secs and which possesses the dimensionality $data_{train} \in R^{1000x1}$. As mentioned in Section~\ref{sec:gkkf_algo_preprocess_data}, the time signal is now split into single overlapping chunks. 
When using a sampling interval $T_{C} = 0.05$ and a chunk length of $w = 1$ the signal is split into 200 unique chunks that all contain 100 data points, where always 20 subsequent chunks overlap. In the next step, the \ac{ft} is computed which returns a data series of 100 complex numbers for every chunk which represents a single observation in the frequency domain. As mentioned before, we only have to keep the first $N$ frequencies but need to split them into their real and imaginary parts which in the end leaves us with a series of 102 data points for every chunk. Altogether, our training set will now have the dimensionality $data_{train} \in R^{200x102}$. As we can see, the ability to work with the traffic data needs to be paid with an increase of the observation dimensionality.

 \label{fig:comp_tf}

\begin{figure}
   \centering
   \includegraphics[width=0.8\textwidth]{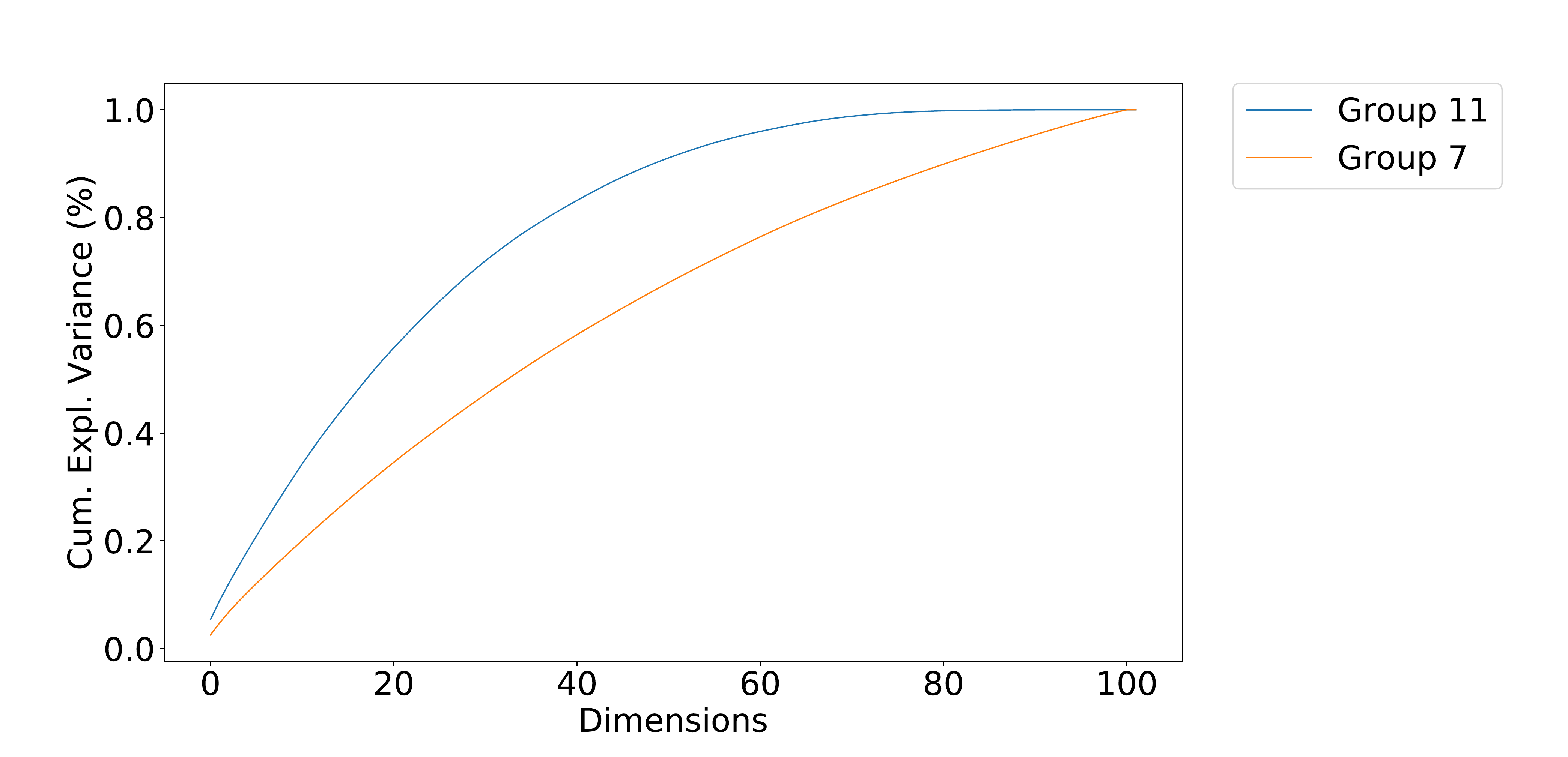}
   \caption{Comparison between lowest and highest cumulative explained variance per dimension}
   \label{fig:pca_dim_var}
 \end{figure}

As discussed in Section~\ref{sec:gkkf_algo} a \ac{pca} was used to reduce the complexity of the training data set and save computation time.
The number of dimensions by which the data sets can be reduced with a \ac{pca} differs between the groups of recurrent flows. 
Figure~\ref{fig:pca_dim_var} shows the cumulated explained variance over the principal components for two flow groups which represent the groups with the best and the worst cumulated explained variances respectively.
As the plot reveals, fewer dimensions are needed for flow group 11 than for flow group 7 to explain the same amount of variance. The variance values were obtained by selecting all flows except one flow of each group as separate training sets and performing a \ac{pca} on them. 
The influence of the reduction on the flow structure in the time domain when reducing the dimensionality from 102 to 80 dimensions is acceptable. For flow group 7, a reduction of 20 dimensions leaves us with 89.9\% of explained variance. 
The loss of information is still acceptable, even though it is clearly visible in the plot. For flow group 11, the remaining 80 dimensions are still enough to explain 99.8\% of the variance. 
However, the plot demonstrates that for this flow group a larger percentage of information was contained in the omitted 20 dimensions since the heights of the flow peaks differ more from the original height. Therefore, the results match the insights already obtained from Figure~\ref{fig:pca_dim_var}. 
Considering also the results for other flow groups (see Table~\ref{table:pred_exp} column \textit{``PCA cum. expl. Variance (80 dim.)''}), it has to be concluded that a \ac{pca} is useful for reducing the dimensionality of the utilized traffic data, a reduction of around 20 dimensions lead to a noteworthy reduction of complexity for further computation. 

As mentioned in Section \ref{sec:introduction}, the training sets used during the experiments contain observations of 
kbit values which not represent the true (hidden) state of the underlying system.
Our training sets only contain observations, however, as noted in Section~\ref{sec:learningPhase}, a window of observations can be used as an internal state representation. 
By computing the \ac{ft} over chunks of the signal we are actually already building a window because the state is represented by the frequencies of a time signal chunk that consists of multiple observations. 
However, to extend the representation of the state during the prediction experiments, it was formed using the \acp{ft} of multiple chunks with increasing length. 
One possibility would e.g. be to use the \ac{ft} of the next \secs and combine it with the \acp{ft} computed over the next 2 and 3\secs. 
This allows us to include some more \textit{long}-term behavior of the flow in the state representation. In such a scenario, an observation would be represented by the \ac{ft} over the next second of the time signal.

\begin{table}[ht]

\small
\setlength{\tabcolsep}{3mm}
\centering
\begin{tabular}{|c||c|c|c|c||c|c|c|c|}
\cline{6-9}
\multicolumn{5}{c|}{} & \multicolumn{4}{c|}{\bf{prediction error}}\\
\hline
\bf{\specialcell{group\\ID}} & \bf{\# flows} & \bf{\specialcell{PCA\\cum. expl.\\Variance\\(80 dim.)}} & \bf{\specialcell{constant\\error}} &  \bf{\specialcell{optimal\\chunk len.}} & \bf{\specialcell{chunk len.\\= 1\secs}} & \bf{\specialcell{chunk len.\\= optimal}} & \bf{ARIMA} & \bf{GARCH} \\
\hline
1 & 4 & 95.9\%&	-69.7\% & 0.15\secs &\cellcolor{tbl_y} -43.1\% &\cellcolor{tbl_g} -2.0\% & -96.3\% & -97.9\% \\
\hline
2 & 13 & 91.7\%& -60.1\% & 0.4\secs &\cellcolor{tbl_y} -36.3\% &\cellcolor{tbl_g} -8.4\% & -100\% & -98.6\% \\
\hline
3 & 6 &	98.7\% & -63.5\%  & 0.55\secs &\cellcolor{tbl_g} 14.7\% &\cellcolor{tbl_g} 5.6\% & -100\% & -96.6\% \\
\hline
4 & 24 & 94.9\% & -57.6\% & 0.2\secs &\cellcolor{tbl_r} -62.3\% &\cellcolor{tbl_r} -22.8\% & -99.9\% & -98.2\% \\
\hline
5 & 5 & 98.1\% & -55.9\% & 0.2\secs &\cellcolor{tbl_r} -53.1\%&\cellcolor{tbl_g}  -12.1\% & -91.8\% & -95.5\% \\
\hline
6 & 6 & 90.6\% & -46.9\% & 0.15\secs &\cellcolor{tbl_r} -52.8\% &\cellcolor{tbl_g} -13.2\% & -97.7\% & -97.5\% \\
\hline
7 & 3 & 89.9\% & -60.0\% & 0.25\secs &\cellcolor{tbl_g} -4.5\% &\cellcolor{tbl_g} 2.5\% & -94.1\% & -97.4\% \\
\hline
8 & 10 & 97.0\% & -79.6\% & 0.3\secs &\cellcolor{tbl_y} -28.5\% &\cellcolor{tbl_g} -4.3\% & -100\% & -97.6\% \\
\hline
9 & 4 & 99.5\% & -66.7\% & 1.95\secs &\cellcolor{tbl_r} -28.4\% &\cellcolor{tbl_g} -4.1\% & -99.9\% & -95.7\% \\
\hline
10 & 4 & 98.5\% & -63.2\% & 0.5\secs &\cellcolor{tbl_y} -46.7\% &\cellcolor{tbl_g} -0.7\% & -100\% & -99.3\% \\
\hline
11 & 5 & 99.8\% & -54.3\% & 0.25\secs &\cellcolor{tbl_y} -26.8\% &\cellcolor{tbl_g} -0.8\% & -99.9\% & -99.0\% \\
\hline
12 & 18 & 95.9\% & -60.4\% & 0.95\secs &\cellcolor{tbl_y} -44.2\% &\cellcolor{tbl_g} -10.2\% & -91.9\% & 4102.7\% \\
\hline
13 & 5 & 94.3\% & -45.5\% & 0.55\secs &\cellcolor{tbl_g} -11.6\% &\cellcolor{tbl_g} -1.6\% & -91.7\% & 98.7\% \\
\hline
14 & 3 & 98.1\% & -62.5\% & 0.35\secs &\cellcolor{tbl_r} -63.8\% &\cellcolor{tbl_g} -10.2\% & -99.7\% & 7490990.2\% \\
\hline
15 & 11 & 90.6\% & -52.2\% & 0.25\secs &\cellcolor{tbl_r} -57.9\% &\cellcolor{tbl_g} -17.0\% & -97.7\% & -98.9\% \\
\hline
16 & 13 & 93.2\% & -37.5\% & 0.55\secs &\cellcolor{tbl_r} -70.6\% &\cellcolor{tbl_r} -40.2\% & -99.0\% & -95.2\% \\
\hline
17 & 3 & 91.4\% & -26.4\% &  0.95\secs &\cellcolor{tbl_r} -93.2\% &\cellcolor{tbl_r} -45.4\% & -90.8\% & -99.3\% \\
\hline
18 & 4 & 98.6\% & -35.7\% & 0.45\secs &\cellcolor{tbl_r} 151.9\% &\cellcolor{tbl_g} 4.7\% & -95.8\% & 13550.9\% \\
\hline
19 & 7 & 93.4\% & -70.0\% & 0.15\secs &\cellcolor{tbl_r} -53.5\% &\cellcolor{tbl_g} -7.3\% & -100\% & 603\% \\
\hline
20 & 3 & 94.8\% & -62.5\% & 0.95\secs &\cellcolor{tbl_r} -53.7\% &\cellcolor{tbl_g} 2.0\% & -77.3\% & 1533.6\% \\
\hline
\end{tabular}
\caption{Results of prediction experiments. While our \ac{fgkkf} achieves 10.9\% prediction error in average over all flow groups, ARIMA and GARCH exhibit at best 77.3\% and  95.2\% prediction error respectively.}
\label{table:pred_exp}
\end{table}

\subsection{Results}
\label{sec:expResults}
Finally, the traffic flows that originated from recurring connections are clustered in groups dependent on the characteristic in frequency space.
In further traffic prediction experiments the clustered flow groups were used to test if the flow structure of single groups could be learned and used to predict unseen flows.  

\paragraph{Prediction accuracy}
20 groups with the highest numbers of flows were selected. 
The results are shown in Table~\ref{table:pred_exp}, 
where the column \textit{``\# of flows''} holds the information about how many flows each group contained. 
On average, one group consisted of 8 unique flows. 
The performance of the \gls{fgkkf} was tested on every flow group separately. 
During an experiment, 
always one flow was selected as the test set. 
The rest of the flows were used for constructing the training set (cf. \other).
As noted previously, the data set representing a flow consists of a time series of kbit values. For the prediction experiments described in this section, a sampling interval $T_{S} = 0.01$ was used for producing the data series and a chunk sampling interval of $T_{C} = 0.05$ for splitting them into smaller chunks.

In Section~\ref{sec:exp_data} we showed the results of our traffic data analysis which revealed that \gls{http} traffic flows often consist of short traffic peaks that in turn can be divided into a rising phase and a phase with a mainly constant load level \cite{microte}. 
Therefore, the more important parts of the flows are the peak rises, which is why we explicitly concentrate our predictions on these parts. We tested the ability of the \gls{fgkkf} to predict the peak rises after only a few observations from the beginning of the peak. The number of observations differs between the flows and ranges between 3 and 60 steps in the frequency domain which corresponds to an observed time interval of 0.15\secs to 3\secs. When the observation stops and the true prediction begins, all flows still possess a low traffic load which then increases rapidly by around 150\% on average during the next \secs. According to Benson et al.'s definition~\cite{microte} all used traffic flows are \textit{un}predictable at the point of the last observation. 

 \begin{figure}
 	\centering
 	\includegraphics[width=18cm]{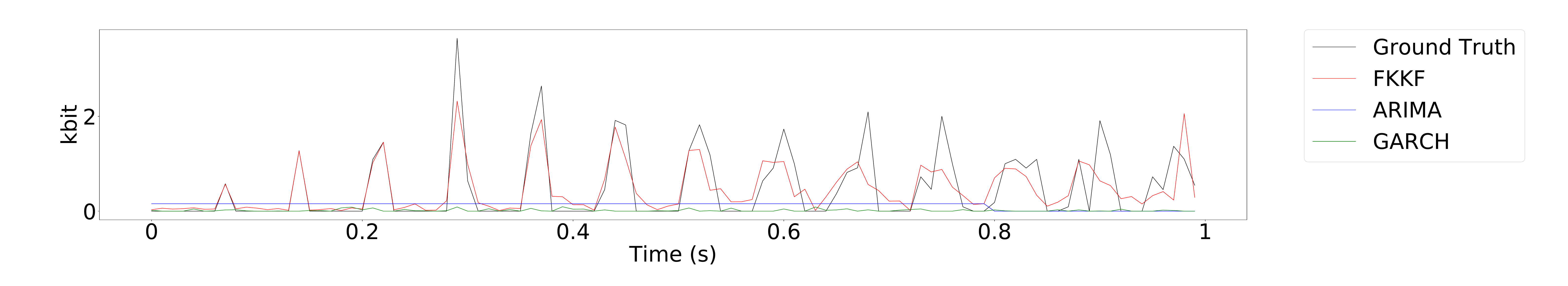}
 	\caption{Comparison of \ac{fgkkf}, ARIMA and GARCH on flow group 3}
 	\label{fig:pred_err_no_3}
 \end{figure}
 \begin{figure}
 	\centering
 	\includegraphics[width=18cm]{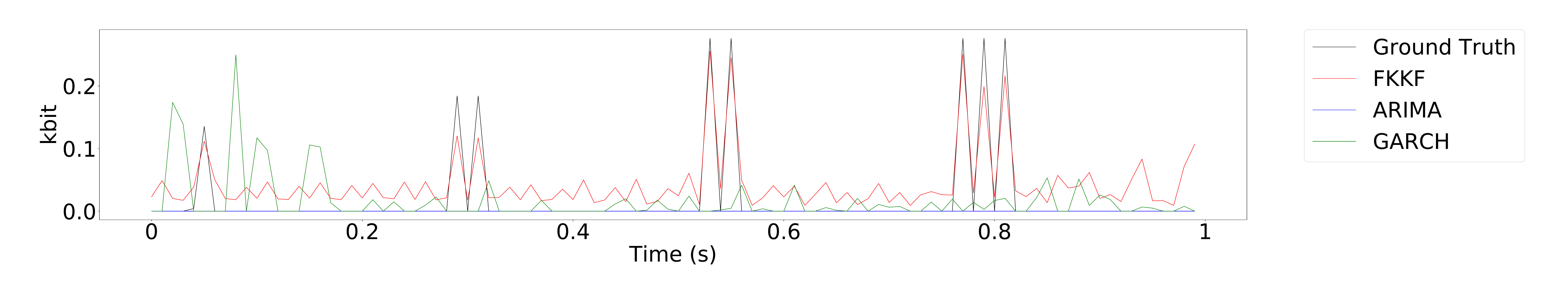}
 	\caption{Comparison of \ac{fgkkf}, ARIMA and GARCH on flow group 19}
 	\label{fig:pred_err_no_19}
 \end{figure}
 \label{fig:pred_err_no}

The \gls{fgkkf} was always used for predicting a time period of 1\secs, resulting in 20 prediction steps in the frequency domain. The highest single impulse value predicted by the \gls{fgkkf} during this period was then compared to the actual highest impulse value in the same period. 
The difference between both values divided by the actual highest value represents the applied error metric for the prediction.  
The error should always be compared to the information that was already given in the observations.  
Therefore, another error metric was calculated for every flow group which helps to assess the quality of the prediction.  
In Table~\ref{table:pred_exp} the metric is shown in the \textit{``constant error''} column. 
It was calculated like the prediction error only that the highest impulse value which emerged during the observation phase was used as the prediction,  just as if we were \textit{predicting} a constant load.  
Since we only observe the beginning of the peak rise,  
the picked impulse value will always be smaller than the actual one during the next \secs.  
Consequently, 
the actual value will always be underestimated which is why all values in the column are negative.

During the first set of experiments a chunk length $w = 1$ was used for all flow groups.  
The internal state representation was then built by combining the \glspl{ft} of the next \secs, the next 2\secs and the next 3\secs of the signal. 
An observation was represented by the \gls{ft} of the next \secs. The prediction errors obtained for all flow groups are shown in column \textit{``chunk len. = 1\secs''}. The performance of the \gls{fgkkf} for predicting a particular flow group is indicated by a red, yellow, or green coloring of the associated table cell which corresponds to a bad, moderate or good prediction performance respectively. Since it is difficult to assess the true prediction quality solely based on error metrics, a two-step approach was used.

In the first step, the computed prediction error and constant error values were utilized. Whenever the prediction error was above 50\% or equally bad as the corresponding constant error value, the quality of the prediction was considered insufficient and the flow group marked in red. When the prediction error was lower than 50\% and clearly smaller than the constant error, the group was marked in yellow. Only if a prediction error of less than 20\% was achieved, the prediction was classified as \textit{good} and marked in green. 
To illustrate the prediction results Figure~\ref{fig:pred_err_no} shows two prediction exemplary results of group 3 (Figure~\ref{fig:pred_err_no_3}) and group 19 (Figure~\ref{fig:pred_err_no_19}) for all considered approaches and the ground truth data.
As expected, ARIMA and GARCH are not able to learn the hidden state of the system at every chunk length and consider the peaks as outliers (ARIMA) and predict zeros or very low values, or start to oscillate arbitrarily (GARCH) while the \ac{fgkkf} algorithm is able to predict the peaks accurately.

 \begin{figure}
 	\centering
 	\includegraphics[width=1\textwidth]{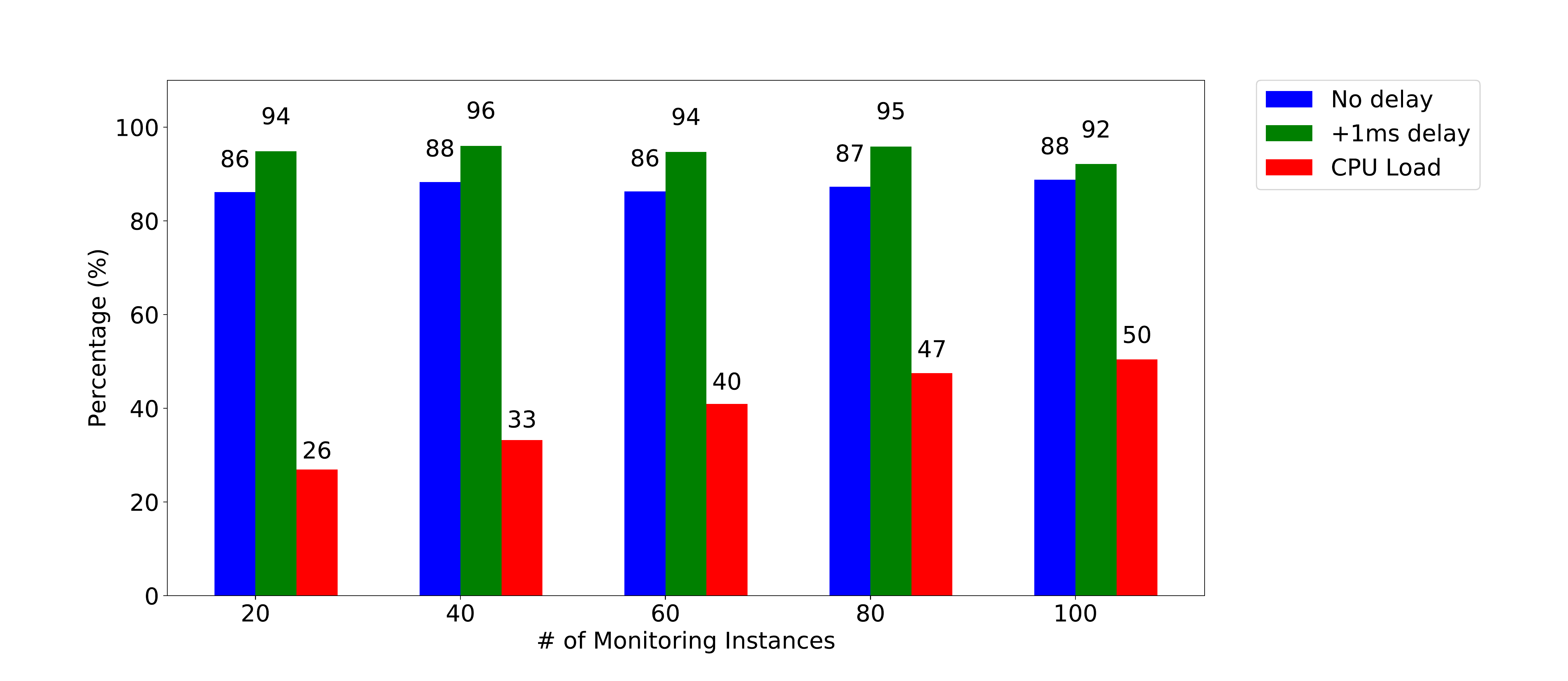}
 	\label{fig:MoSeSEval}
 \caption{Accuracy and CPU load of the monitoring system with different number of monitoring instances and a polling interval of 10ms}
 \label{fig:MoSeS}
 \end{figure}

 \paragraph{Overhead}
 Being able to run our \ac{fgkkf} algorithm on commodity data center switch hardware is as important as 
achieving good prediction results. For the system evaluation we used a \mbox{AS5512-54X}~ 10G~\footnote{{https://www.edge-core.com/\_upload/images/AS5512-54X\_DS\_R03\_20180614.pdf}} switch from Edgecore  with OpenSwitch.
As discussed in Section~\ref{sec:sys_design} the \ac{fgkkf} required a new monitoring system to poll the statistic of individual flows periodically at  exact points in time because inaccurately measured data can highly influence the observed trajectory, leading to very inaccurate predictions.
Therefore the developed monitoring system was optimized for accurate polling. 
Figure~\ref{fig:MoSeS} shows the result of the monitoring polling accuracy with an interval of 10\msecs. 
On the $x$ axis different numbers of monitoring instances where tested. 
The blue bars show the accuracy without any delay.
The green bars visualize the accuracy with 1\msecs delay which is still in an acceptable range.
The red bars show the CPU load of one core with the given number of monitoring instances simultaneously.
The tested switch has a quad core CPU for the management system, which
 means that by monitoring 100 individual flows only 50\% of one core is occupied.
As mentioned in Section~\ref{sec:sys_design} the monitoring instance sends a copy of the monitored data to an instance which is responsible for reclassifying and relearning of a flow, if the prediction results of an individual flow are not accurate enough. 
This step is not time critical and can be computed on other devices with more computational resources (cf.~\simult).
Therefore, it is not considered in the evaluation.

As shown in Figure~\ref{fig:switch} on top of the monitoring instances the \ac{fgkkf} instances are predicting a given flow.
One \ac{fgkkf} prediction needs, on only one core, between 0.1\msecs and 10\msecs (on average 3.65\msecs) for  running 1000 predictions in different flow groups. The variance is due to cache misses and context switches of the \ac{os} and computational optimization of the learned model. 
As shown in Table~\ref{table:pred_exp} the optimal chunk length varies between different flow groups. 
The average of optimal chunk length is 0.49\secs, which is the interval between two predictions. 
By taking the average processing time of the \ac{fgkkf} algorithm into account, more than 200 flows can be predicted on one switch (cf. \simult). 
\section{Conclusions}
\label{sec:conclusion}

In this paper we proposed an
approach for predicting traffic flows focused on the short term prediction of peak structures of individual flows because,  as shown during the analysis of real network traffic, 
this is the only type of prediction that truly considers the existing bursty nature of the network traffic that is a main cause of network congestion. 
In the conducted traffic prediction experiments, 
our \ac{fgkkf} was used to predict the peak rises in flow groups consisting of single flows from recurrent socket-to-socket connections, 
with observations only being given to the \ac{fgkkf} while the flows' traffic load was still low. In 17 of the observed 20 groups, a prediction of the highest traffic impulse could be achieved with an average error of 6.43\%.

Decentralized data processing (at the network devices) should also be considered in the development of any prospective \ac{te} system that aims at using prediction for rerouting network traffic. 
One idea is to realize several heterogeneous systems
which will be simultaneously learning to optimize a global reward signal, e.g., to minimize the amount of network congestion for the given traffic profile. 
Due to the exploration actions of all agents, the global reward signal might be almost uncorrelated with the actions and observations of single agents. 
Therefore, the agents can combine the \ac{fgkkf}'s prediction with additional information from a global controller instance. 
 
Our \ac{fgkkf} can be also used for mitigating other problems in the network than for dealing with congestion. 
Another interesting avenue for future research we are investigating is network security
Here we are using our \ac{fgkkf} to learn and predict patterns of (distributed) denial-of-service attacks.


\end{spacing}
\end{document}